\documentclass[useAMS,usenatbib]{mn2e}
\usepackage[a4paper,margin=2cm]{geometry}
\usepackage[pdftex]{graphicx}
\pdfoutput=1
\usepackage{color}
\usepackage{aas_macros}
\usepackage{lastpage}
\usepackage{comment}
\usepackage{amsmath}
\usepackage{type1cm}
\usepackage{eso-pic}
\usepackage{url}

\def\lesssim{\mathrel{\hbox{\rlap{\hbox{\lower4pt\hbox{$\sim$}}}\hbox{$<$}}}}
\def\gtrsim{\mathrel{\hbox{\rlap{\hbox{\lower4pt\hbox{$\sim$}}}\hbox{$>$}}}}
\newcommand{\ltaraw}{$\; \buildrel < \over \sim \;$}
\newcommand{\lta}{\lower.5ex\hbox{\ltaraw}}
\newcommand{\gtaraw}{$\; \buildrel > \over \sim \;$}
\newcommand{\gta}{\lower.5ex\hbox{\gtaraw}}
\def\lsim{\mathrel{\rlap{\lower4pt\hbox{\hskip1pt$\sim$}}
    \raise1pt\hbox{$<$}}}                

\newcommand{\ie}{{\it i.e.}}
\newcommand{\eg}{{\it e.g.}}

\newcommand{\LCDM}{$\Lambda$CDM}
\newcommand{\Vmax}{$V_{\rm max}$}
\newcommand{\VSF}{$V_{\rm SF}$}
\newcommand{\VSFz}{$V_{\rm SF}{(z)}$}

\newcommand{\Rx}{$\mathcal{R}_{\rm x}$}
\newcommand{\Rz}{$\mathcal{R}_{\rm z}$}
\newcommand{\Rh}{$\mathcal{R}_{\rm h}$}
\newcommand{\Rt}{$\mathcal{R}_{\rm t}$}
\newcommand{\Rv}{$\mathcal{R}_{\rm x}(s,V_{\rm max})$}
\newcommand{\Rs}{$\mathcal{R}_{\rm x}(s)$}

\newcommand{\bx}{$b_{\rm x}$}
\newcommand{\bz}{$b_{\rm z}$}

\newcommand{\sx}{$s_{\rm x}$}

\newcommand{\Dfb}{$\Delta f_b$}
\newcommand{\Dfs}{$\Delta f_s$}
\newcommand{\kmax}{$k_{\rm max}$}

\def\apjl{ApJ}%
\def\apjs{ApJS}%
\def\ao{Appl.~Opt.}%
\def\aap{A\&A}%
\title[GiggleZ: Scale Dependant Bias]{The Gigaparsec WiggleZ Simulations: \\ Characterising scale dependant bias and associated systematics in growth of structure measurements}

\author[Poole et al.]{\parbox[t]{\textwidth}{
    Gregory B.\ Poole$^{1,2}$\footnotemark,
    Chris Blake$^1$,
    Felipe A. {Mar{\'{\i}}n}$^1$,
    Chris Power$^3$, \\
    Simon J. Mutch$^2$,
    Darren J. Croton$^1$, 
    Matthew Colless$^4$, 
    Warrick Couch$^5$, \\
    Michael J.\ Drinkwater$^6$, 
    Karl Glazebrook$^1$} \\ \\ 
  $^1$ Centre for Astrophysics \& Supercomputing, Swinburne University of Technology, P.O. Box 218, Hawthorn, VIC 3122, Australia \\ 
  $^2$ School of Physics, University of Melbourne, Parksville, VIC 3010, Australia \\
  $^3$ ICRAR, University of Western Australia, 35 Stirling Highway, Crawley, Western Australia 6009, Australia \\
  $^4$ Research School of Astronomy \& Astrophysics, Australian National University, Weston Creek, ACT 2600, Australia \\ 
  $^5$ Australian Astronomical Observatory, P.O. Box 915, North Ryde, NSW 1670, Australia \\ 
  $^6$ School of Mathematics and Physics, University of Queensland, Brisbane, QLD 4072, Australia}

\date{draft version \today}
\pagerange{\pageref{firstpage}--\pageref{lastpage}} \pubyear{2012}

\begin{document}

\label{firstpage}

\maketitle

\begin{abstract}
We present the Gigaparsec WiggleZ (GiggleZ) simulation suite and use this resource to characterise galaxy bias and its scale dependence for a range of redshifts and halo masses in a standard \LCDM\ cosmology.  Under the ansatz that bias converges to a scale independent form at large scales, we develop an 8-parameter phenomenological model which fully expresses the mass and redshift dependence of bias and its scale dependence in real or redshift space.  This is then used to illustrate how scale-dependent bias can systematically skew measurements of the growth-rate of cosmic structure obtained from redshift-space distortion measurements.  When data is fit only to scales $k_{\rm max}{\le}0.1$ $[h^{-1} \rm{Mpc}]^{-1}$, we find that these effects are significant only for large biases ($b{\gtrsim}3$) at large redshifts ($z{\gtrsim}1$).  However, when smaller scales are incorporated ($k_{\rm max}{\lsim}0.2$ $[h^{-1} \rm{Mpc}]^{-1}$) to increase measurement precision, the combination of reduced statistical uncertainties and increased scale dependent bias can result in highly significant systematics for most large halos across all redshifts.  We identify several new interesting aspects of bias, including a significant large-scale bias boost for small halos at low-redshifts due to substructure effects ($\sim$20\% for Milky Way-like systems) and a nearly redshift-independent halo mass (corresponding to a redshift-space bias of ${\sim}1.5$) for which halo bias has little-or-no scale dependence on scales greater than $3$ $[h^{-1} {\rm Mpc}]$.  This suggests an optimal strategy of targeting bias ${\sim}{1.5}$ systems for clustering studies which are dominated more by systematic uncertainties in how observed halo (or galaxy) distributions map to their underlying mass distribution than by observational statistical precision, such as cosmological measurements of neutrino masses.  Code for generating our fitting formula is publicly available at \url{http://gbpoole.github.io/Poole_2014a_code/}.
\end{abstract}


\section{Introduction}\label{sec-intro} 
\renewcommand{\thefootnote}{\fnsymbol{footnote}}
\setcounter{footnote}{1}
\footnotetext{E-mail: gpoole@unimelb.edu.au}

Maps of the distribution of galaxies across enormous cosmic volumes -- as determined from galaxy redshift surveys -- have become extremely rich resources for a variety of powerful examinations of cosmological models.  These include (but are certainly not limited to) precise standard ruler measurements of the cosmic expansion history using harmonic features induced by ``Baryon Acoustic Oscillations'' (BAOs) in the Universe's matter density field and measurements of the growth rate of cosmic structure as probed by the imprints of the cosmic peculiar velocity field on redshift-derived (\ie\ redshift-space) distributions of galaxies.  Our ability to perform these and other cosmological examinations using redshift surveys is based upon our ability to connect observed galaxy distributions to our highly developed and robust models of the distribution of matter in the early Universe, its evolution with redshift and the dependence of both on background cosmology.  Of course, the success of this endeavour rests completely on our ability to relate observed galaxy distributions to their underlying matter distributions; a relationship generally referred to as ``galaxy bias''.  However, it has long been understood observationally that galaxy bias has a complicated dependancy on galaxy luminosity, colour and morphology \citep{Loveday:1995p2458,Hermit:1996p2457} with modern studies still continuing to refine this understanding \citep[\eg][see \citealt{Baugh:2013p2490} for a review]{Norberg:2001p2477,Zehavi:2005p2449,Ross:2007p2534,Swanson:2008p2533,Cresswell:2009p2532}.

In this article we seek to characterise this relationship.  We aim to build a phenomenological parameterisation of halo bias and it's scale dependance across a range of masses and redshifts which we can use this to ascertain when and to what degree scale dependant bias becomes important for observational measurements of the growth rate of cosmological structure.  Several ancillary outcomes will also result including a correction to large-scale bias estimates for substructure effects or for systems exhibiting strong scale-dependant bias, useful for other studies intimately linked to the redshift evolution of bias.

As with most aspects of large scale structure, a great deal of theoretical insight can be obtained through excursion set analyses.  The earliest successful theory of this type was that of \citet[][subsequently extended by \citealt{Bardeen:1986p2507}]{Kaiser:1984p2506} who illustrated how the two-point clustering statistics of collapsed cosmological objects becomes enhanced if associated with early overdensities in the cosmological matter field.  The first model to build explicitly upon the popular framework of \citet{Press:1974p2491} and its extensions (EPS) was that of \citet[][MW]{Mo:1996p1904} which was subsequently confronted by the numerical investigation of \citet{Jing:1998p1968} who identified significant discrepancies in this model's treatment of lower-mass systems.  These discrepancies were traced to incorrect assumptions about the form of the halo mass function in MW by \citet{Sheth:1999p2125} who were able to build a successful analytic model constructed from mass functions calibrated by numerical simulations, thus establishing an intimate link between the mass-dependent clustering bias of a halo population and its associated mass function.  This was soon followed by \citet[][SMT]{Sheth:2001p2201} who added an account of the dynamics of ellipsoidal collapse to the traditional EPS approach through the adoption of a mass-dependence for the spherical collapse overdensity, leading to significant improvements in the excursion set results for both mass functions and the mass dependence of large-scale bias \citep[although, see][for a recent challenge to this interpretation]{Borzyszkowski:2014p2544}.  

Generally, two approaches to the analysis of halo bias exist: Eulerian approaches (which dominate the literature) focus on the contemporaneous relationship of halo and matter clustering and Lagrangian approaches which relate the evolving clustering of halos to their \emph{initial} linear-regime matter field.  Interesting challenges to the conclusions of Eulerian studies have emerged from Lagrangian studies.  For example, \citet{Porciani:1999p1951} utilised simulations to show that the low-mass bias modifications of \citet[][mentioned above]{Jing:1998p1968} to the analytic model of MW reflects conditions embedded in the initial state of the simulations, and not exclusively subsequent non-linear processes.  Such findings motivate a careful examination of traditional excursion set descriptions of halo formation; a conclusion echoed by \citet{Jing:1999p2185} and subsequently built upon by several studies including \citet{Ludlow:2011p2509} and \citet{Elia:2012p2492}.

While analytic progress continues to be made \citep[\eg][who employ a Non-Markovian extension and a stochastic collapse barrier within the framework of traditional EPS approaches to obtain improved mass function and bias models]{Ma:2011p2451}, the work of SMT makes it clear that treatment of the detailed structure of collapsing cosmological fields are important to obtaining accurate estimates of volume-averaged clustering statistics.  As a result, most significant progress has been driven of late by improved calibrations of analytic models using N-body simulations \citep[\eg][]{Seljak:2004p2217,Tinker:2005p2354}.  This effort has culminated in \citet[][TRK]{Tinker:2010} who examine a more generalised form of the SMT model and perform a careful numerical calibration of its parameters.  Recent studies have validated the TRK model \citep[\citet{Papageorgiou:2012p2274}; see][for a similarly successful model]{Basilakos:2001p2280,Basilakos:2008p2299} which we will use as our main comparison for the large-scale bias calculations which anchor the scale-dependent bias analysis in this work.

While large-scale galaxy bias has received a great deal of study, relatively few inquiries have been made into its scale dependence.  Early examinations \citep[\eg][]{Sheth:1999p2536,CasasMiranda:2002p2537,Zehavi:2004p2535,Seo:2005p2356} have discussed some general expectations and presented evidence of scale-dependant bias in observed datasets but the work of \citet{Tinker:2005p2354} is the first to present a general model.  Subsequently, in their clustering analysis of the 2dF Galaxy Redshift Survey, \citet{Cole:2005p1578} introduced the ``Q-model''; a phenomenological Fourier-space model which has subsequently found applications in the analysis of SLOAN LRGs \citep[\eg][]{Padmanabhan:2007p1128}.  Employing arguments based on the halo model, other Fourier-space accounts of scale-dependent bias include the model of \citet[][subsequently extended by \citealt{Huff:2007p2355}]{Schulz:2006p2447} and \citet{Smith:2007p2450} who clearly illustrate the existence of scale dependent bias and a dependence on halo mass and galaxy type.  Furthermore, \citet{Pollack:2013p2508} have recently explored scale dependent bias within standard perturbation theory finding that the non-linear processes giving rise to such effects are not sufficiently described in popular second-order local Eulerian schemes.  Lastly, \citet{Paranjape:2013p2547} have shown how to identify and remove scale dependant bias effects using simulated halo catalogs resulting in large-scale biases which are in good agreement to those of TRK.

Observationally, an important additional complication arises.  Positions for very large ensembles of galaxies are generally not determined through direct distance measurements but are rather inferred from redshifts.  Distributions measured in this way are said to be constructed in ``redshift-space'' and the presence of peculiar velocities imprinted upon the background Hubble-flow by accelerations from local density gradients is known to induce significant bias effects in this space.  The classic treatment by \citet[][K87 henceforth]{Kaiser:1987p1760} predicts that coherent bulk flows on large scales induce a ``Kaiser-boost''; a significant increase in clustering bias over that which would be inferred in real-space due to halo assembly effects alone.  On large scales, this model has been validated by numerical simulations \citep[\eg][]{Montesano:2010p2505} but on small scales -- where incoherent motions such as those giving rise to the ``Fingers of God'' effect can lead to a \emph{suppression} of bias -- significant scale-dependence to these redshift-space effects have been identified \citep[\eg][]{Seljak:2001p2480}.

The primary consequence of scale dependent bias is that it introduces a source of systematic uncertainty to cosmology constraints at a level which is now important for ongoing and future surveys.  While several investigations have found that the consequences for constraints based on the scale of the BAO peak should not be significant \citep[\eg][a conclusion supported by our investigations]{Eisenstein:2007p2482,Crocce:2008p2481,Angulo:2008p2483,Smith:2008p2484} there is concern that constraints sensitive to the full \emph{shape} of scale-dependent clustering statistics (such as power spectra or correlation functions) will be more susceptible, particularly when pushed to smaller scales.  Two notable such cases include measurements of neutrino masses from the cosmological power spectrum \citep[\eg][]{RiemerSrensen:2012p1754} and measurements of the growth rate of cosmic structure \citep[\eg][]{Blake:2011b} which we focus on in this study.

Beyond the issue of systematic bias, several interesting physical processes can lead to scale dependent bias providing new opportunities for the study of other physics.  These include the induction of scale-dependent bias from departures from non-Gaussianity in the early universe \citep{Dalal:2008p1750,Slosar:2008p1751,Taruya:2008p2459} or from subtle environmental effects induced by the physics of galaxy formation \citep{Coles:2007p2454,Barkana:2011p2453}.  Firmly establishing an accurate and robust theory in the absence of these effects will be essential for their search in observational datasets.

In this work, we take a distinctly different approach from past studies, performing a straight-forward phenomenological characterisation of Eulerian bias in configuration space.  Surprisingly little theoretical investigation of scale dependent bias within this framework has been performed in the recent literature despite the fact that it's the space in which most observational analysis is performed.  As noted by \citet{Huff:2007p2355} \citep[also see][]{Guzik:2007p2448}, configuration space offers an important advantage over Fourier space: a lower amplitude of scale dependent bias.  Interpreted within the framework of the halo model, they note that this is due to the fact that most scale dependent bias is a product of the different scales on which matter and galactic halos transition from the 1-halo regime to the 2-halo regime.  This occurs on relatively small scales as far as most cosmological studies are concerned, thus isolating its effects in configuration space.  In Fourier space, such broad-spectrum features become spread across a wider range of scales transferring signal from the small scales on which the phenomena occurs, to larger scales where most of the clustering signal resides.

We use the Gigaparsec WiggleZ (GiggleZ) Simulation Suite for this study.  GiggleZ was constructed to support the science program of the WiggleZ Dark Energy Survey \citep{Drinkwater:2010} -- a large redshift survey of UV-selected galaxies conducted with the multi-object AAOmega fibre spectrograph at the 3.9-m Australian Astronomical Telescope -- and has been used in several WiggleZ-related publications to date \citep[\eg][]{Blake:2011a,RiemerSrensen:2012p1754,Contreras:2013p2488,Marin:2013p2487,Blake:2013p2486}.  We take this opportunity to present details related to the construction of the GiggleZ simulation program and subsequently present a simple and direct model of the mass and redshift dependence of both large-scale and scale-dependent bias of dark matter halos.  We examine for the first time the effects of substructure on models of galaxy bias of this form, finding significant ($\sim$20\%) effects on low-bias systems at low redshift.  We then use this model to build upon previous studies of systematic biases in growth of structure measurements \citep{Okumura:2011p2540,Jennings:2011p2541,Contreras:2013p2488}, calculating the potential magnitude of systematic errors induced in the absence of corrections for scale-dependent bias effects.

In Section \ref{sec-simulations} we present the GiggleZ simulation suite; the simulations involved, our approach to initialising, running and analysing them, and the results of a convergence study run to determine the optimal integration properties of our adopted simulation code.  In Section \ref{sec-analysis} we present our scale dependent bias model, stepping through the justifications for each of our chosen parameterisations.  In Section \ref{sec-cosmology_biases} we present the consequences of scale dependent bias for growth of structure measurements. Lastly, we summarise and discuss our conclusions in Section \ref{sec-summary}.

Our choice of fiducial cosmology throughout will be a standard spatially-flat WMAP-5 $\Lambda$CDM cosmology \citep{Komatsu:2009}: ({$\Omega_\Lambda$}, {$\Omega_M$}, {$\Omega_b$}, {$h$}, {$\sigma_8$}, {$n$})=(0.727, 0.273, 0.0456, 0.705, 0.812, 0.960).

\section{Simulations}\label{sec-simulations}
\begin{table*}
\begin{minipage}{170mm}
\begin{center}
\begin{tabular}{ccccccc}
\hline
Simulation	&
$L$ $[{h^{-1}}\rm{Mpc}]$& 
$N_p$	& 
$m_p$  [10${^9}$ M$_\odot$/h] &
${\rm n}_{\rm snap}$ &
$\Delta t$ [Myrs]  &
$\epsilon$ $[{h^{-1}}\rm{kpc}]$\\
\hline
GiggleZ-main	&  1000	& 2160$^3$	&  7.52	& 100 & 115	& 9.3 \\
GiggleZ-LR	&   125	& 135$^3$	&  60.13	& 931 & 15	& 18.5 \\
GiggleZ-NR	&   125	& 270$^3$	&  7.52	& 931 & 15	& 9.3 \\
GiggleZ-MR	&   125	& 540$^3$	&  0.95	& 467 & 30	& 4.6 \\
GiggleZ-HR	&   125	& 1080$^3$	&  0.12	& 235 & 60	& 2.3 \\
\hline
\end{tabular}
\caption{Box sizes ($L$), particle counts ($N_p$), particle mass ($m_p$), number of snapshots (${\rm n}_{\rm snap}$), approximate snapshot temporal resolution ($\Delta t$) and gravitational softening length ($\epsilon$) for the GiggleZ simulations. \label{table-simulation_parameters}}
\end{center}
\end{minipage}
\end{table*}

\begin{figure*}
\begin{center}
\begin{minipage}{175mm}
\includegraphics[width=175mm]{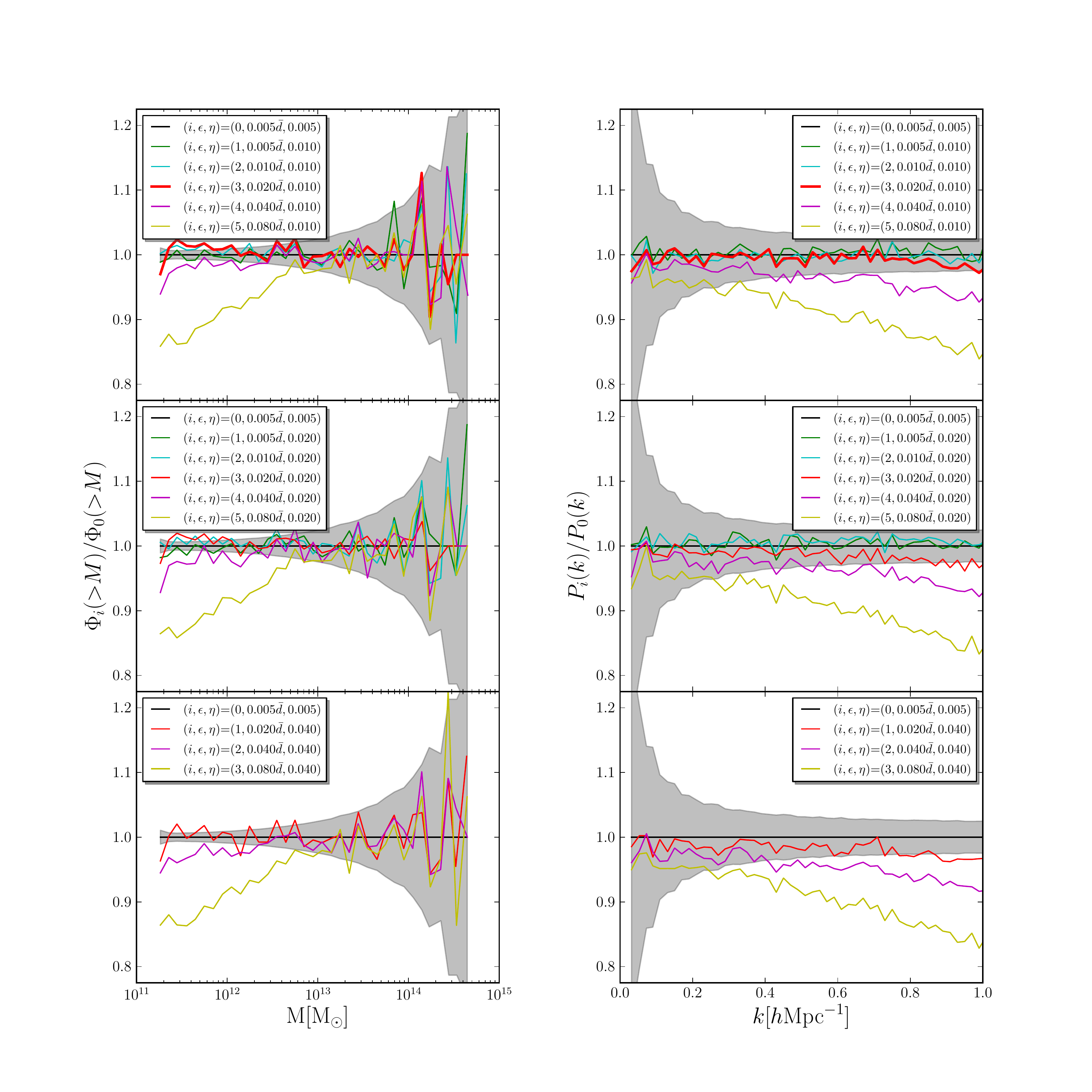}
\caption[Convergence tests]{A plot comparing the effects of variations in gravitational softening parameter ($\epsilon$) and time step integration accuracy ($\eta$) on the mass function ($\Phi_i$; left) and power spectrum of halos more massive than 10$^{12}$ M$_\odot$ ($P_i$; right) of a $(L,N)$=(250 [${h^{-1}}\rm{Mpc}$], $540^3$) simulation (\ie\ the same mass resolution as the GiggleZ-main and GiggleZ-NR runs).  In each case, we normalise the mass function and power spectrum to the case ($\epsilon$,$\eta$)=($0.005\bar{d}$,$0.005$).  Based on these results, we selected ($\epsilon$,$\eta$)=($0.02\bar{d}$,$0.01$) for all GiggleZ simulations (where $\bar{d}$ is the mean interparticle spacing of the simulation).  Runs with $\epsilon{=}0.02\bar{d}$ are labeled in red with the $\eta{=}0.01$ case additionally highlighted with a thick line (top panels).  Grey shaded regions indicate the magnitude of the Poisson statistical uncertainty of the ($\epsilon$,$\eta$)=($0.005\bar{d}$,$0.005$) case used as reference in all cases. \label{fig-convergence}} 
\end{minipage}
\end{center}
\end{figure*}

The GiggleZ simulation suite consists of 5 simulations: a large GiggleZ-main run consisting of 2160$^3$ particles distributed in a periodic box $1$ $[{h^{-1}}\rm{Gpc}]$ on-a-side, and 4 simulations of an identical $125$ $[{h^{-1}}\rm{Mpc}]$ on-a-side control volume spanning a factor of 512 in mass resolution with snapshot temporal resolutions as fine as 15 Myrs.  The basic specifications for these 5 runs are listed in Table \ref{table-simulation_parameters}.   The large scale of the GiggleZ-main simulation was motivated by the unprecedented combination of large volume and low halo mass of the low-bias UV-selected galaxies targeted by WiggleZ.  Such observational programs present a demanding challenge for theoretical support of clustering studies, leading us to create (at the time) one of the highest-resolution gigaparsec-scale cosmological simulations available, comparable to modern simulation programs such as the Multi-dark BigBolshoi Simulation \citep{Prada:2012p2538}.  The control-volume simulations were designed to conduct systematic studies of the resolution requirements for semi-analytic galaxy formation studies.  In this paper we focus on the GiggleZ-main simulation only.  A companion paper will present the control volume simulations in detail where they are used to present our method of merger tree construction and their convergence properties.

We have run our simulations with GADGET-2 \citep{Springel:2005b}, a Tree-Particle Mesh (TreePM) code well suited to large distributed memory systems.  We have modified the publicly available version to conserve RAM in dark matter only simulations by removing all support for hydrodynamics, 'FLEXSTEP' time stepping and variable particle masses (along with all associated memory allocations).  All simulations were run on the Green Machine at Swinburne University, with the largest run consuming all the resources of 124 nodes, each housing dual quad core Intel Clovertown 64-bit processors (for a total of 992) with 16GB of RAM.

\subsection{Initial Conditions}

To initialise our simulations we use the Parallel N-body Initial Conditions (PaNICs) code developed at Swinburne for this project.  PaNICs follows the approach of \citet{Bertschinger:2001} to construct a displacement field which, when applied to a uniform distribution of particles, yields a distribution with our desired power spectrum.  This power spectrum was generated using CAMB \citep{Lewis:2000} with our standard spatially-flat WMAP-5 $\Lambda$CDM cosmology given above.  This power spectrum was normalised for a starting redshift $z_{init}$=49 for the GiggleZ-main run and $z_{init}$=499 for the control volume simulations.  These starting redshifts ensure that initial particle displacements are smaller than the grid cell size of the displacement field for all simulations, a condition advocated by \citet{Lukic:2007}.  This high starting redshift may introduce some numerical noise for the lower resolution control volume runs affecting detailed halo structure, but should have a negligible effect on the mass accretion histories which will be the main focus of their use.  This was verified for the GiggleZ-NR mass resolution during our convergence testing in which we performed a run with $z_{init}$=49 and found no significant effect on the simulation's halo power spectrum or mass function.

For the GiggleZ-main simulation, the displacement field was computed on a 4320$^3$ grid while the control volume simulations used a common displacement field computed on a 2160$^3$ grid.  Uniform distributions in all cases were computed from integral periodic tilings of a 135$^3$ glass configuration \citep[see][for more details]{White:1994} generated using GADGET.

Particle velocities were computed from the PaNICs displacement field using the Zeldovich approximation \citep{ZelDovich:1970,Buchert:1992}.  Higher-order corrections to this calculation \citep[\eg][]{Scoccimarro:1998p2493,Crocce:2006p2494} could not be implemented in a timely fashion for this project, but will certainly be incorporated in future projects.

\subsection{Halo finding}

The majority of the analysis in this study will utilise the bound dark matter halos which emerge from our simulations.  To extract these structures we use the well tested code SUBFIND of \citet{Springel:2001}.  This code first starts by finding friends-of-friends (FoF) structures for which we use the standard linking length criterion of 0.2$\bar{d}$ (where $\bar{d}=L/\sqrt[3]{N_p}$ denotes the mean interparticle spacing of the simulation).  It subsequently identifies bound substructures within these FoF groups as locally overdense collections of particles, removing unbound particles through an unbinding procedure.  

This procedure leads to two classes of halo: FoF groups and substructure halos.  In the work which follows, we perform our analyses on both classes of halo separately.  Since FoF groups are more closely related to the overdensity peaks forming the basis of Extended Press-Schechter analyses, results derived from study of these objects should form a better comparison to models developed within that framework.  However, observed galaxy populations are more closely related to our substructure halos and results derived from analyses of this class of halo should be more straight-forwardly related to observed galaxy distributions.  Later in Section \ref{sec-trends} we will find that there are interesting differences between the bias properties of the two.

\subsection{Convergence tests}

Being principally responsible for the accuracy and run-time of our simulations, we carefully considered the settings of two GADGET parameters in particular when setting-up our calculations: the gravitational softening ($\epsilon$; we will express this in units of $\bar{d}$ henceforth) and the dimensionless parameter controlling the accuracy of the timestep criterion ($\eta$; referred to as $\rm{ErrTolIntAccuracy}$ in the GADGET manual).

We ran a grid of $(L,N)$=(250 [${h^{-1}}\rm{Mpc}$], $540^3$) simulations (\ie\ the same mass resolution as the GiggleZ-main and GiggleZ-NR run), varying combinations of these parameters over the ranges $\epsilon$=0.005$\bar{d}$ to 0.08$\bar{d}$ and $\eta$=0.005 to 0.04.  Since our primary science interests in WiggleZ involve studies of L* galaxy formation and clustering on 100 [${h^{-1}}\rm{Mpc}$] scales, we seek convergence based on the substructure halo mass function and substructure halo power spectrum of halos in the range M${>}10^{12}$ [$h^{-1}$ M$_{\odot}$].

The results are presented in Fig. \ref{fig-convergence}.  Expected trends are realised: larger softenings in particular have a strong impact on small scales (\ie\ low-mass and high-$k$).    Furthermore, we find that the power spectrum is a more stringent condition in these tests than the mass function.  When $P(k)$ is converged, the mass function is converged.  Using the power spectrum at $z$=0 as our metric of fitness, we can immediately rule out softenings $\epsilon{>}0.04\bar{d}$ by demanding that deviations from our fiducial $P(k)$ remain less than 5\% over the range $k$=0.1 to 1 $[h^{-1} \rm{Mpc}]^{-1}$.

There is a degeneracy in these tests between $\epsilon$ and $\eta$: moderate increases in $\epsilon$ can be compensated for by decreasing $\eta$.  Reducing $\eta$ has a significant impact on the run-time of the simulation however, placing practical constraints on how far it can be lowered.  Taken in combination, we use these constraints to settle upon the combination ($\epsilon$,$\eta$)=(0.02$\bar{d}$,0.01) for all runs in this project.  From these experiments, we expect the mass function to be accurate to ${\sim}{2}$\% on M* scales.  We expect the power spectrum to be accurate to ${\sim}{2}$\% over the range $k{=}\left[0.1,1\right]$ $[h^{-1} \rm{Mpc}]^{-1}$. 

\begin{figure}
\includegraphics[width=80mm]{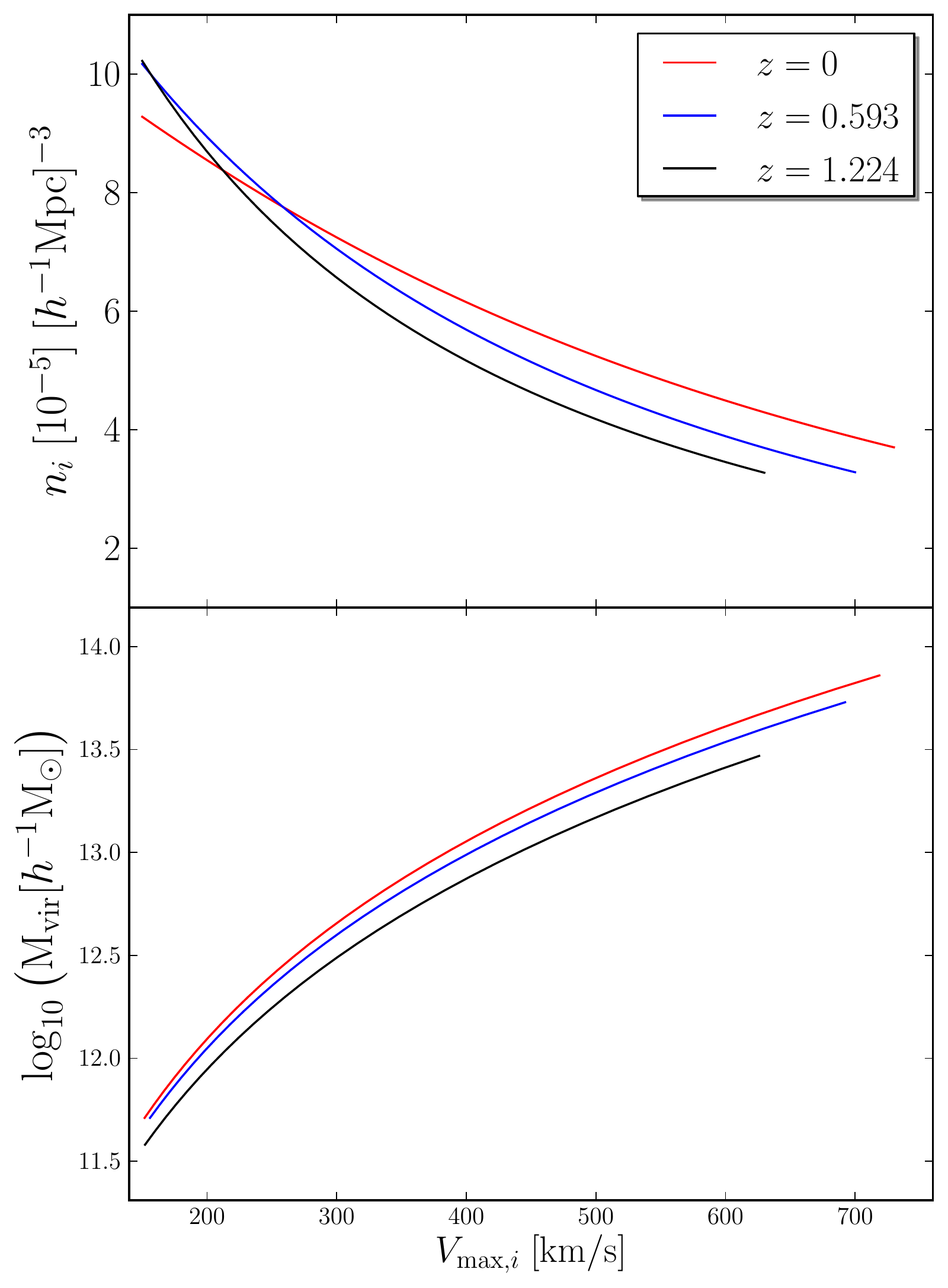}
\caption[Grouping number densities]{A plot presenting the number densities ($n_i$) adopted for the halo 'groupings' and the relationship between \Vmax\ and $M_{\rm vir}$ used for all analysis in this work.  Three redshifts evenly spanning the range of this study ($z{\lsim}1.2$) are depicted.  Number densities are chosen such that they scale inversely with large scale bias (as estimated from the model of \citealt{Tinker:2010}, TRK) and the linear growth factor (\ie\ $n_i{\propto}1/\left(b_{TRK}D\right)$) normalised such that $n_i\left(z_i{=}0.6{,}b_{TRK}{=}1\right){=}10^5$ [(h/Gpc)$^3$].\label{fig-n}} 
\end{figure}

\subsection{Halo groupings}\label{sec-groupings}

For this study, we are interested in the mass and redshift dependence of halo clustering properties.  To facilitate our analysis, we have assembled a number of 'groupings' of both our FoF and substructure halos for a set of seven redshifts from $z{=}0$ to $z{\sim}1.2$ in steps of $dz{\sim}0.2$.  In each case we have rank-ordered the structures by their maximum circular velocities (denoted $V_{max}$) and selected contiguous groupings of ${n_i}\left(z_i,V_{{\rm max},i}\right)$ systems (yielding grouping number densities of $n_i$ per $\left[{h^{-1}}\rm{Gpc}\right]^3$) for each 'i'th grouping.  This is done such that $V_{{\rm max},i}$ are median values for their respective groupings, starting at $150$ km/s for $i{=}0$ and extending upwards in steps of $10$ km/s until we run out of massive halos (at a value of $V_{{\rm max},i}$ which declines with redshift).  We use \Vmax\ as our metric of halo mass to render our results less sensitive to peculiarities of our chosen halo finder and to increase reproducibility.  Furthermore, subhalo abundance matching has suggested that \Vmax\ may more directly parameterise the stellar mass of galaxies \citep{Reddick:2013p2519}, potentially improving the degree to which our \Vmax-selected subhalo groupings represent the clustering characteristics of stellar-mass selected galaxy samples.  See Figure \ref{fig-n} for an illustration of the relationship between \Vmax\ and $M_{\rm vir}$.

We set $n_i$ for each grouping to yield correlation functions of roughly equivalent signal-to-noise despite the growth of structure moderated by the linear growth factor (denoted $D$ and given by $D{=}\delta(z)/\delta(0)$, where $\delta(z)$ is the evolving matter density contrast) and the mass-dependent bias which we estimate using the \citet[][TRK]{Tinker:2010} model\footnote{Throughout this paper we will convert the overdensity parameterising this model to an effective \Vmax\ assuming standard \citet{Navarro:1997p2543} scaling properties and the mass-concentration relation of \citet{MunozCuartas:2011p2542}.} (denoted $b_{TRK}$).  This approach has the added benefit of naturally reducing our bin size as mass and redshift increase, adapting to regimes where halos densities are low and clustering properties are rapidly evolving.  More specifically, we choose groupings for which ${b_{TRK}}\left(V_{{\rm max},i}{,}z_i\right){=}1$ to have $n_i{=}10^{5}$ halos at $z{=}0.6$ and scale $n_i$ for other cases by $1/\left(b_{TRK}D\right)$.  The resulting values of $n_i$ used for this study are illustrated in Figure \ref{fig-n}.

To add redshift-space distortion effects to our catalogs we assume a flat-sky approximation, taking the positions of each halo grouping and adding a 1D displacement in the $x$-direction ($\delta{x}$) given by:
\begin{equation}\label{eqn-z_space_distortions}
\delta{x}{=}\frac{v_x h}{a(z)H(z)}
\end{equation}

\noindent where $v_x$ is the $x$-component of the physical centre-of-mass velocity of the halo, $a(z)$ is the cosmological expansion factor and $H(z)$ is the redshift-dependant Hubble parameter.

\section{Analysis}\label{sec-analysis}
\begin{figure*}
\begin{minipage}{170mm}
\begin{center}
\includegraphics[width=120mm]{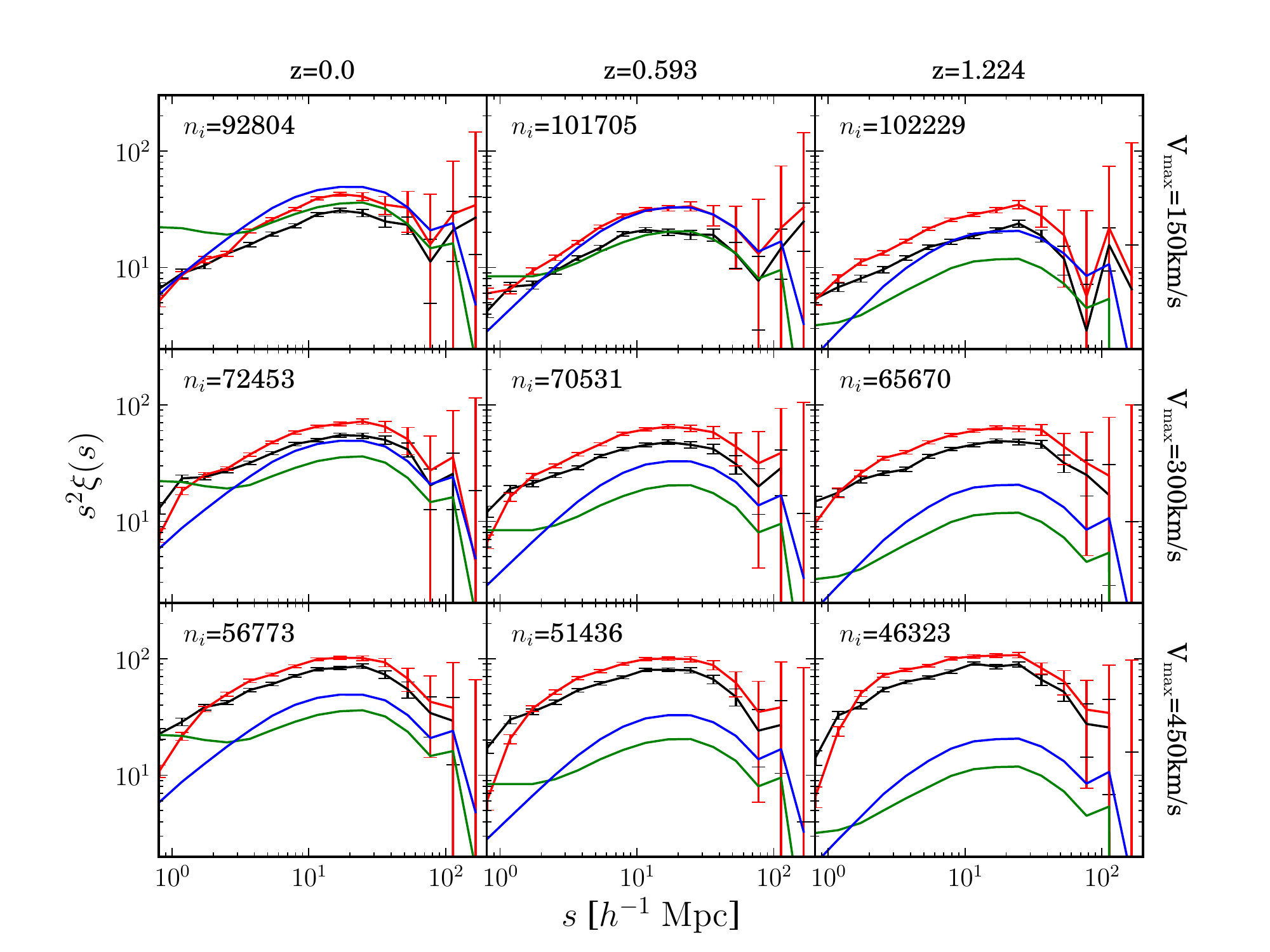}
\caption[correlation functions] {Two-point correlation functions ($\xi(s)$; plotted as $s^2\xi(s)$ with an extra factor of $s^2$ to increase figure clarity) for the total matter and for populations of dark matter halos at three halo masses and three redshifts.  Green and blue lines denote $\xi(s)$ in real and redshift-space for the dark matter particles at each redshift respectively.  Black and red lines with error bars denote $\xi(s)$ in real and redshift-space for the dark matter halos respectively.  In all cases, the number halos involved in the represented FoF halo groupings ($n_i$) is given.  Uncertainties are computed from jack-knife subsamples using a regular $6^3$ grid.  Values across the top denote the redshift represented by each column while values along the right indicate the halo mass (expressed in terms of maximum halo circular velocity, \Vmax) represented by each row.}\label{fig-correlation_functions}
\end{center}
\end{minipage}
\end{figure*}

\begin{figure*}
\begin{minipage}{170mm}
\begin{center}
\includegraphics[width=120mm]{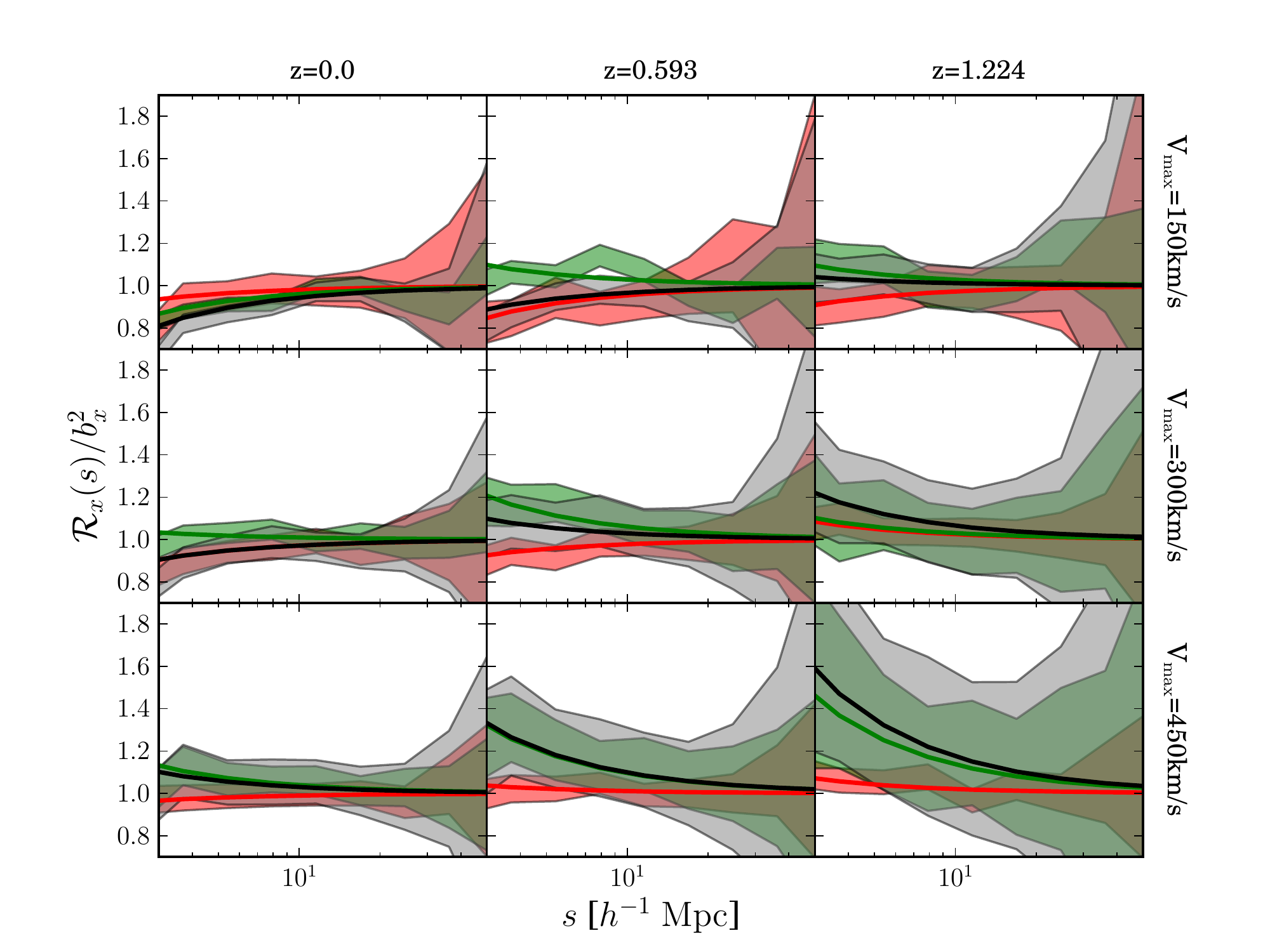}
\caption[scale dependence ratios] {The scale dependence of three ratios taken between total matter and halo correlation functions at the same three halo masses and three redshifts depicted in Figure \ref{fig-correlation_functions}.  Red denotes the ratio of the redshift-space halo correlation function to the real-space halo correlation function (\Rz; expressing the square of redshift-space boost effects on the total redshift-space bias), green the ratio of the real-space halo correlation function to the real-space total matter correlation function (\Rh; expressing the square of the real-space halo bias) and grey the ratio of the redshift-space halo correlation function to the real-space total matter correlation function (\Rt; expressing the square of the total redshift-space bias).  All ratios have been computed using their jack-knife subsamples to minimise cosmic variance, with shaded regions indicating $68$\% confidence intervals.  Thick solid lines indicate the best fit of Eqn. \ref{eqn-R_of_s} to each dataset assuming $\eta{=}1$.  In all cases, the square of the large-scale bias ($b_{\rm x}^2$) has been normalised-out such that all curves converge to a value of $1$ at large values of $s$.  Values across the top denote the redshift represented by each column while values along the right indicate the halo mass (expressed in terms of maximum halo circular velocity, \Vmax) represented by each row.} \label{fig-scale_dependence_ratios}
\end{center}
\end{minipage}
\end{figure*}

For the analysis presented in this paper, we will use the 2-point correlation function as our measure of clustering strength and its scale dependence.  The method of \citet{Landy:1993p1846} is used throughout and is applied to all of the halo groupings described in Section \ref{sec-groupings} as well as to randomly sampled subsets of $10^6$ particles from each relevant snapshot of our simulations.  This method requires a large number of randomly distributed points and we use $250000$ points for halo analysis and $5{\times}10^6$ for matter field analysis, ensuring that there are at least 5 times more random points than data points in all cases.

Examples of our computed correlation functions are presented in Figure \ref{fig-correlation_functions} where we show results at three redshifts evenly spanning the range of our study ($z{=}0$,$0.593$ and $1.224$) for distributions of matter and for FoF halo groupings of three masses (\Vmax${=}150$, $300$, and $450$ [km/s]) in the GiggleZ-main simulation.  Expected trends of increasing clustering amplitude with halo mass and increased redshift-space clustering (particularly on scales less than ${\sim}2$ [$h^{-1}$ Mpc] where "halo exclusion" effects become significant) are apparent.  

\subsection{Computing scale dependent bias and motivating its general form}\label{sec-computing_b_r}

\noindent Throughout the analysis which follows, we will focus on three correlation function ratios which capture separate contributions to halo bias and its scale dependence.  These ratios will be between the redshift-space halo correlation function and the real-space halo correlation function  (\Rz; said to  express the redshift-space -- or Kaiser, after the model of \citealt{Kaiser:1987p1760} -- 'boost' effects on total bias), the ratio of the real-space halo correlation function to the real-space dark matter correlation function (\Rh; said to express the real-space halo bias) and the ratio of the redshift-space halo correlation function to the real-space dark matter correlation function (\Rt; said to express the total or redshift-space bias).  Throughout this work we will refer to these ratios in a general form as \Rx\ where $x$=`z',`h' or `t' denoting the Kaiser boost, halo or redshift-space bias ratios respectively.  Conceptually, \Rt${=}$\Rh${\times}$\Rz, although we fit to each ratio individually and do not enforce this relation.

In all cases, \Rx$(s)$ profiles and uncertainties are computed from the median and (potentially asymmetric) distribution of 216 jack-knife subsamples evaluated using a regular $6^3$ grid.  This choice for the number of jack-knife regions was motivated by the work presented in \citet{Contreras:2013p2488} where we found that results are insensitive to the number of regions used for the regimes studied here.  Furthermore, we concentrate only on scales larger than $3$ [$h^{-1}$ ${\rm Mpc}$] for two reasons: we find that the behaviour of \Rx$(s)$ on scales less than this is complicated \citep[with a character similar to that presented in figure 4 of][]{Zehavi:2004p2535} and difficult to parameterise and because it is on scales less than this where the morphology-density relation of observed galaxy populations becomes significant \citep{Hansen:2009p2521,Haines:2009p2522,vonderLinden:2010p2523,Lu:2012p2524,Wetzel:2012p2525,Rasmussen:2012p2526,Bahe:2013p2528}, greatly complicating the use of these scales for realistic galaxy populations.  

Examples of each ratio for cases spanning the range of redshift and halo mass addressed by this study are shown in Figure \ref{fig-scale_dependence_ratios} (for all plots henceforth, the same colour scheme is used: green to represent real-space halo bias, red to represent Kaiser boost effects and black to represent total redshift-space bias).  Several general trends are immediately obvious from this plot.  At large scales, the limited volume of our simulation results in a rapid increase in the variance of each \Rx\ profile as scales begin to exceed 20-30 [$h^{-1}$ Mpc]. Within these admittedly large uncertainties, there is little evidence of scale-dependent bias effects beyond these scales, as we expect from the results of previous studies.  At smaller scales where our simulation is adequate for quantifying \Rs, we see clear evidence of scale dependence increasing in magnitude with halo mass and redshift for \Rh\ and \Rt\ while trends are more mild and less discernible for \Rz.  Furthermore, in some regimes we find that \Rx\ can be enhanced on small scales relative to large scales (generally the case for \Rh\ and \Rt) or suppressed on small scales.

Therefore, taking as an ansatz that \Rx\ converges to a constant value at large scales \citep[although, see][for evidence of slight scale-dependent effects on scales ${>}140~\lbrack h^{-1}~\rm{Mpc} \rbrack$, albeit at a level insignificant to this study]{Angulo:2014p2548}, this figure motivates us to assume the following form for \Rx:

\begin{alignat}{3}\label{eqn-R_of_s}
			& \mathcal{R}_{\rm x}	&=&~b^2_{\rm x}\left(1+\mathcal{S}\left( s/s_x\right)^{-\eta}\right)\notag\\
\text{where }~	& \mathcal{S}	&=&~{\pm} 1 
\end{alignat}

\noindent This is a four parameter model (applicable on scales $s{>}3 [h^{-1} \rm{Mpc}]$) where $b_x$ quantifies the large-scale bias amplitude, $\eta$ sets the slope of \Rx\ on small scales, $s_x$ is effectively a measurement of the amplitude of scale dependent effects (in the same way that $r_0$ parameterises clustering amplitude when correlation functions take the form $\xi{=}\left(\frac{r}{r_0}\right)^{\gamma}$, particularly for a fixed value of $\eta$ as we will ultimately adopt below) and $\mathcal{S}$ sets whether bias is suppressed by scale dependent effects on small scales (\ie\ the case $\mathcal{S}{=}{-}1$) or enhanced on small scales (\ie\ the case $\mathcal{S}{=}{+}1$).

\subsection{The mass dependence of scale dependent bias}\label{sec-mass_dependence}

\begin{figure*}
\begin{minipage}{170mm}
\begin{center}
\includegraphics[width=115mm]{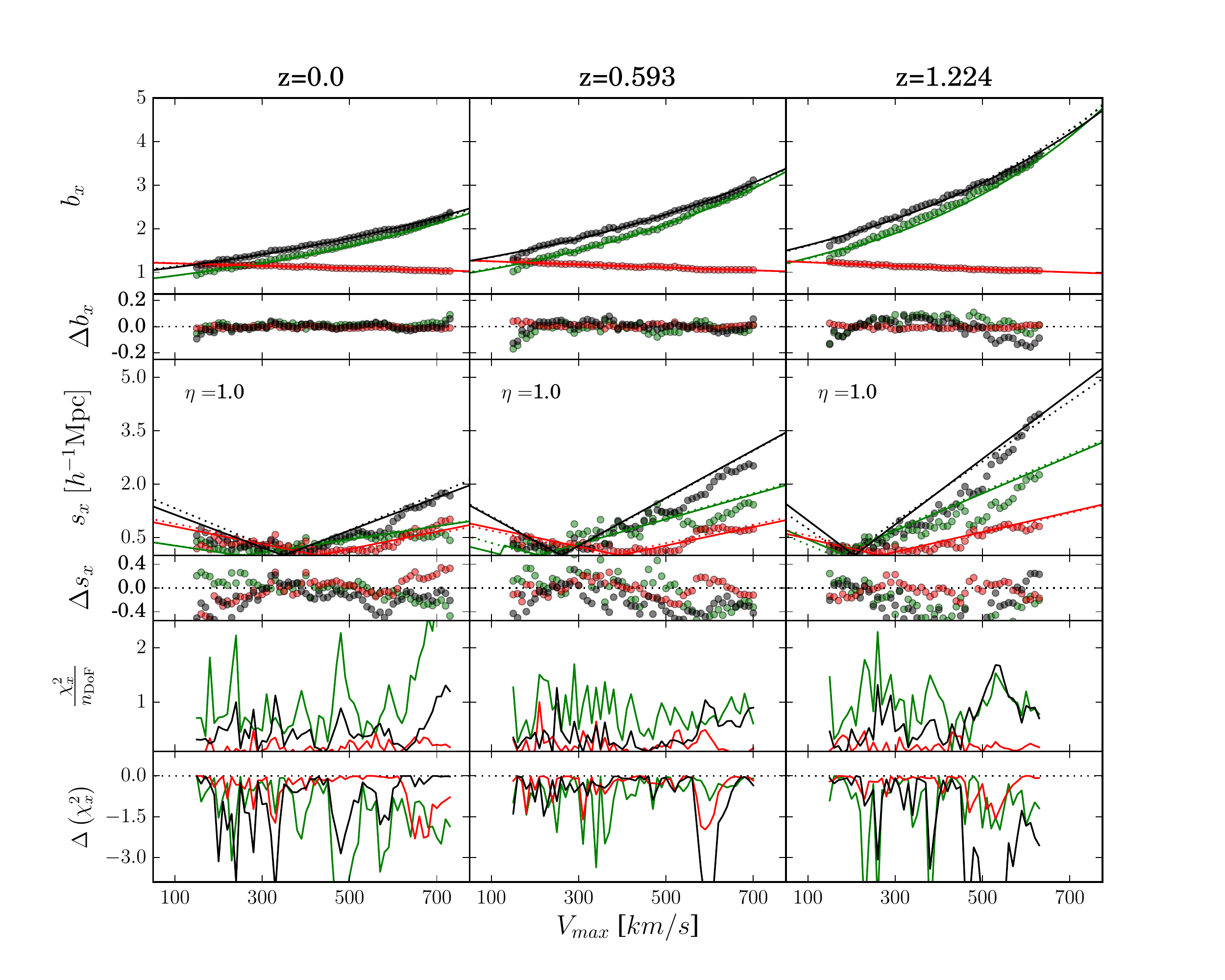}
\caption[Mass dependence] {The results of fitting the scale dependent bias model of Eqn. \ref{eqn-R_of_s} to halo groupings of various masses (expressed in terms of maximum halo circular velocity, \Vmax) at three redshifts spanning the range utilized in this study.  Coloured points indicate fits for redshift-space boost effects (\Rz; red), real-space halo bias (\Rh; green) and total redshift-space bias (\Rt; black).  Solid lines indicate the fit of the halo mass dependence model expressed by Eqn. \ref{eqn-fit_V_dependence} to each of these three datasets (with matching colours).  Dotted lines similarly indicate the results of our full mass-and-redshift dependent bias model expressed by Eqns. \ref{eqn-R_of_s}, \ref{eqn-fit_V_dependence} and \ref{eqn-fit_z_dependence} and Table \ref{table-fit_results}.   Residual differences of each fit from this model are plotted as $\Delta s_x$ and $\Delta b_x$ for the large scale bias and amplitude of scale dependant bias, respectively.  The bottom 2 panels in each column indicate the values of $\chi^2/n_{\rm DoF}$ (where $n_{\rm DoF}{=}6$ in all cases) obtained from the fit assuming $\eta=1$ (second from bottom) and the difference in $\chi^2$ obtained when allowing $\eta$ to vary over the range $0$ to $3$ (bottom;  $n_{\rm DoF}{=}5$ in this case).  Values across the top denote the redshift represented by each column.}\label{fig-fit_V_dependence}
\end{center}
\end{minipage}
\end{figure*}

\begin{figure*}
\begin{minipage}{170mm}
\begin{center}
\includegraphics[width=125mm]{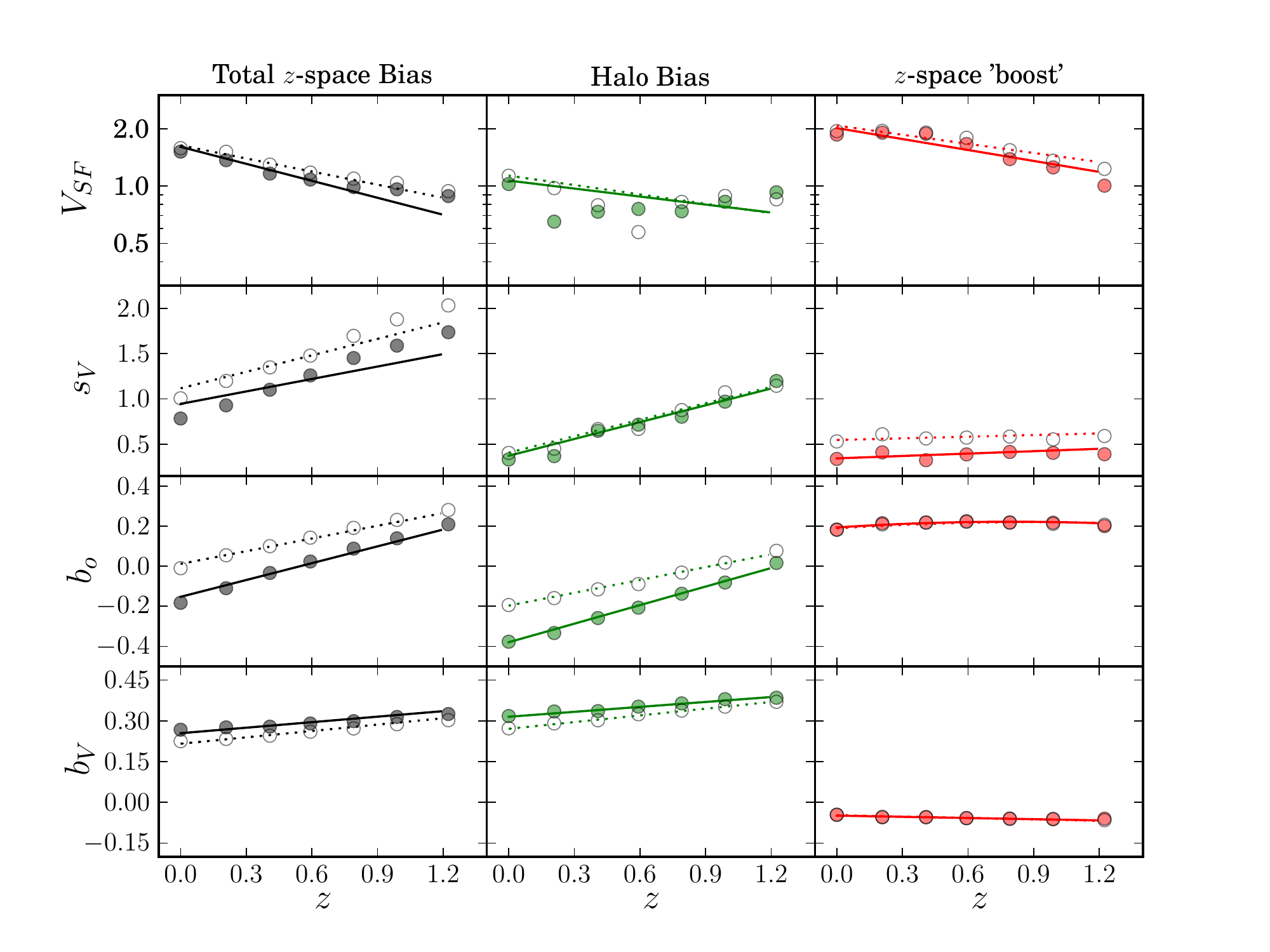}
\caption[Redshift dependence] {The results of fitting our final redshift-and-mass dependent model to the profiles of redshift-space boost effects, real-space halo bias and total $z$-space bias computed for this study.  Individual points denote fits of Eqn. \ref{eqn-fit_V_dependence} to \Rv\ for each ratio type at several redshifts with solid points indicating fits to FoF halos and open points indicating fits to substructure halos.  Solid lines denote our full mass and redshift dependent model expressed by Eqn. \ref{eqn-fit_z_dependence} when fit to FoF halos and dashed lines denote this fit to substructure halos.  Note that the solid lines are not fits to the data points, but rather a single joint MCMC fit to all \Rs\ profiles used in this study.  The agreement validates our chosen parameterisation of the redshift dependence of the parameters in our final model.}\label{fig-fit_z_dependence}
\end{center}
\end{minipage}
\end{figure*}

\begin{table*}
\begin{minipage}{170mm}
\begin{center}
\begin{tabular}{lcccccc}
\hline
Parameter		&
\multicolumn{2}{c}{Real space bias}	& 
\multicolumn{2}{c}{$z$-space boost}	& 
\multicolumn{2}{c}{Total $z$-space bias}	\\ 
 &
FoF Halos & Substructure &
FoF Halos & Substructure &
FoF Halos & Substructure \\
\hline
$V_{\rm SF,x}^{0}$ [220 km/s]				&  0.02819	&  0.05326	&  0.31002	&  0.31731	&  0.20417	 &  0.21287	 \\
$V_{\rm SF,x}^{z}$ [220 km/s]				& -0.13820	& -0.16739	& -0.20264	& -0.15991	& -0.29667	 & -0.22806	 \\
$s_{\rm x}^{\rm V,0}$ [(220 km/s)$^{-1}$]	&  0.36860	&  0.40269	&  0.33423	&  0.53444	&  0.94082	 &  1.11879	 \\
$s_{\rm x}^{\rm V,z}$ [(220 km/s)$^{-1}$]		&  0.61547	&  0.60966	&  0.09233	&  0.07102	&  0.45147	 &  0.61214	 \\
$b_{\rm x}^{\rm 0,0}$					& -0.37936	& -0.19743	&  0.22062	&  0.21988	& -0.15350	 &  0.01198	 \\
$b_{\rm x}^{\rm 0,z}$					&  0.30743	&  0.21382	&  n/a		&  n/a		&  0.27995	 &  0.21127	 \\
$b_{\rm x}^{\rm 0,zz}$					&  n/a		&  n/a		& -0.04419	& -0.03749	&  n/a		 &  n/a		 \\
$z_{\rm b,z}$							&  n/a		&  n/a		&  0.78527	&  0.92920	&  n/a		 &  n/a		 \\
$b_{\rm x}^{\rm V,0}$ [(220 km/s)$^{-1}$]	&  0.31475	&  0.27075	& -0.04805	& -0.04629	&  0.25471	 &  0.21535	 \\
$b_{\rm x}^{\rm V,z}$ [(220 km/s)$^{-1}$]	&  0.06073	&  0.08202	& -0.01454 	& -0.01833	&  0.06761	 &  0.07763	 \\
\hline
\end{tabular}
\caption{Parameters for our full scale dependent bias model, as expressed by Eqns. \ref{eqn-R_of_s}, \ref{eqn-fit_V_dependence} and \ref{eqn-fit_z_dependence}.  Values for both halo types (friends-of-friends halos and substructure) are given. \label{table-fit_results}}
\end{center}
\end{minipage}
\end{table*}

\begin{figure*}
\begin{minipage}{170mm}
\begin{center}
\includegraphics[width=170mm]{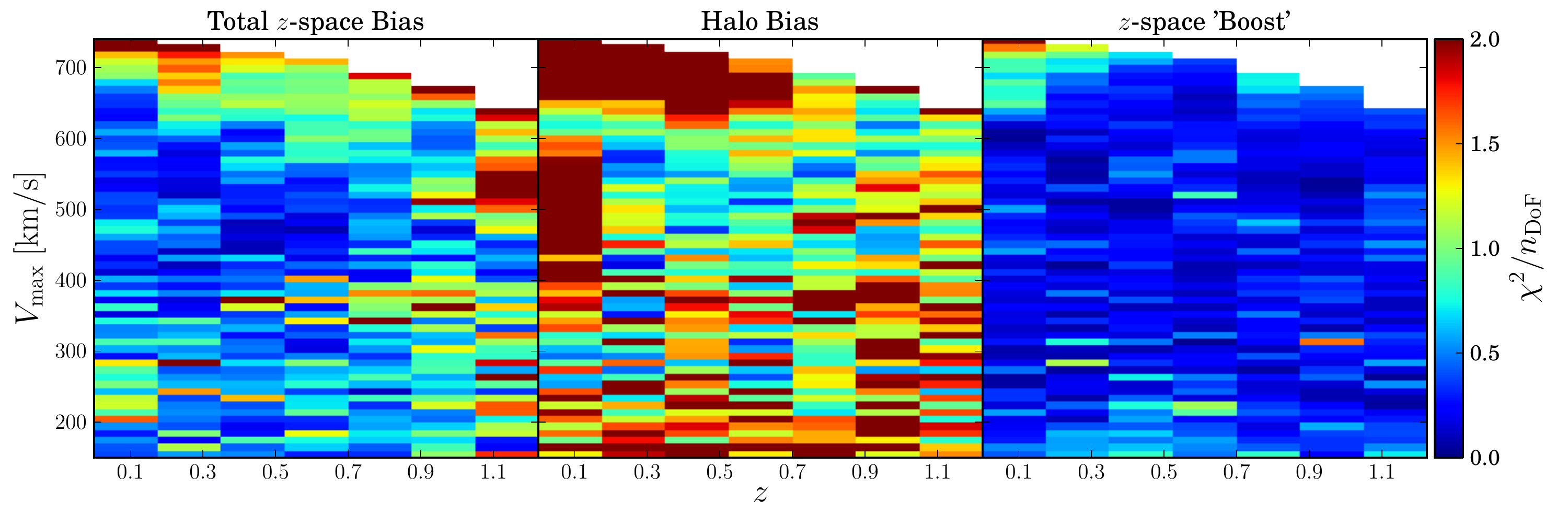}
\caption[chi-squared] {Values of $\chi^2/n_{\rm DoF}$ (where $n_{\rm DoF}{=}6$ in all cases) obtained from fitting our scale-dependent bias model -- $b_{t}{(V_{\rm max},z)}$ on the left, $b_{h}{(V_{\rm max},z)}$ in the middle, $b_{z}{(V_{\rm max},z)}$ on the right -- to the friends-of-friends (FoF) halos of the GiggleZ-main simulation.  These planes represent the full range in \Vmax\ and $z$ over which our model has been constrained, with the white region in the top right being due to a lack of dark matter halos of sufficient density at corresponding masses and redshifts.  \label{fig-fit_chi2}}
\end{center}
\end{minipage}
\end{figure*}

In Figure \ref{fig-fit_V_dependence} we show the results of fitting the model introduced in Equation \ref{eqn-R_of_s} to each scale-dependence ratio, for all of our halo groupings at three redshifts spanning the range of our study.  These preliminary illustrative fits - which at this point are meant only to motivate the parameterisation which follows - are constructed using a simple $\chi^2$-minimisation approach.

When allowing $\eta$ to vary freely between values of $0$ and $3$, we find very little discernible trend for $\eta$ with \Vmax\ and very noisy trends for $s_x$ with $V_{\rm max}$ for all three bias types.  This suggests that the four parameter model of Equation \ref{eqn-R_of_s} is under-constrained by these datasets.  However, when we fix $\eta$ to a value of $1$, clear trends in $s_x(V_{\rm max})$ emerge for all cases, as illustrated in Figure \ref{fig-fit_V_dependence}.  Fixing $\eta$ in this way results in a minimal reduction in the quality of fit, as shown in the bottom panels of this figure where we compare the $\chi^2$ obtained allowing $\eta$ to vary ($n_{\rm DoF}{=}5$) to those obtained when we fix $\eta$ to a value of 1 ($n_{\rm DoF}{=}6$).  This value of $\eta$ was chosen as a compromise in the range of best fit values.  While the results of fits change in detail when other fixed values of $\eta$ are chosen, little change results to the quality of fit or to the conclusions of our study.

For the large scale bias parameters (\bx; illustrated in the top panels of Figure \ref{fig-fit_V_dependence}), expected trends are apparent with real-space and redshift-space halo bias increasing with both mass and redshift.  They follow each other with an offset which decreases with mass but is relatively constant with redshift.  This offset is due to Kaiser boost contributions which we see decline with mass, converging towards a value of 1 (\ie\ no contribution to redshift-space bias from peculiar velocities) as masses increase.  This trend is remarkably constant with redshift as well.

As mentioned above, when we fix $\eta$ to a value of 1, \sx\ effectively quantifies the amplitude of scale dependent bias effects.  For this choice of $\eta$, a value of \sx${=}1$ [$h^{-1}$ Mpc] results in a $15\%$ difference in bias between scales $s{=}3$ [$h^{-1}$ Mpc] and $s{=}\infty$, a value of \sx${=}2$ [$h^{-1}$ Mpc] a $29\%$ difference, etc.

There is a clear pattern illustrated in Figure \ref{fig-fit_V_dependence} of \sx\ decreasing and then increasing roughly linearly with mass about a pivot point which varies with redshift and ratio type.  This is a result of bias effects being suppressed at small scales for small halo masses (\ie\ $\mathcal{S}{=}{-}1$), passing a point at which there is no scale dependence ($s_x{=}0$), and then increasing with enhanced small-scale bias at large values of halo mass (\ie\ $\mathcal{S}{=}{+}1$).  As such, the point of minimum \sx\ for each case indicates a halo mass at which scale dependence of bias disappears.  This behaviour is discernible in Figure \ref{fig-scale_dependence_ratios}.

Motivated by these results, we choose the following parameterisation for the halo mass dependence of scale dependent bias:

\begin{alignat}{3}\label{eqn-fit_V_dependence}
\log_{10}{b^2_{\rm x}(z,V_{\rm max})}	&=&&~b_{\rm x}^{0}(z)+b_{\rm x}^{V}(z)V_{\rm max}   \notag\\
{s_{\rm x}}(z,V_{\rm max})		&=&&~s_{\rm x}^{V}(z)\left|{V_{\rm max}-V_{\rm SF,x}(z)}\right| \notag\\
\mathcal{S}(z,V_{\rm max})		&=&&
\begin{cases}
         -1 & \text{if $V_{\rm max} < V_{\rm SF,x}$,} \\
         +1 & \text{if $V_{\rm max} \ge V_{\rm SF,x}$}
\end{cases}  
\end{alignat}

\noindent  This represents a 4 parameter model describing the mass dependence of bias and its scale dependence at a fixed redshift.  Two parameters describe a linear \Vmax\ dependence for the logarithmic bias ($b^0_x$ and $b^V_x$), one sets the strength of the mass dependence of scale dependent bias ($s^V_x$) and one sets the mass at which bias becomes scale free at the regime between the suppression (at $V_{\rm max}{<}V_{\rm SF,x}$) and the enhancement of bias at small scales (at $V_{\rm max}{>}V_{\rm SF,x}$). The results of fitting this model to the cases illustrated in Figure \ref{fig-fit_V_dependence} are illustrated with solid lines.  For this and all cases which follow, these fits are applied directly to the \Rx\ profiles and their (possibly asymmetric) uncertainty distribution obtained in the manner described in Section \ref{sec-computing_b_r} (and not to the individual points depicted in Figure \ref{fig-fit_V_dependence} resulting from our $\chi^2$ fits to individual cases) using the MCMC machinery introduced in \citet{Poole:2013p1849}.

\subsection{The redshift dependence of scale dependent bias and the final full model}\label{sec-full_model}

Finally, we now seek a parameterisation of the full mass and redshift dependence of scale dependent bias.  This is achieved by parameterising the redshift dependence of the 4 parameters in the model given by Equation \ref{eqn-fit_V_dependence} for each ratio type.

In Figure \ref{fig-fit_z_dependence} we present a series of fits (in coloured points) of the model presented in Eqn. \ref{eqn-fit_V_dependence} at several redshifts spanning the range of our study for both our FoF (solid points) and substructure halos (open points).  Once again, these are preliminary illustrative fits constructed using a simple $\chi^2$-minimisation approach and are meant only to motivate the parameterisation of our final model.  They are equivalent to the fits shown with solid lines in Figure \ref{fig-fit_V_dependence} but applied to a larger number of redshifts (and to both halo types).  We find that the parameters of our mass-dependence model vary smoothly with redshift, motivating the following form for the redshift dependence of scale-dependent bias:

\begin{alignat}{3}\label{eqn-fit_z_dependence}
\log_{10}{V_{\rm SF,x}(z)}	&=&&~V_{\rm SF,x}^{0}+V_{\rm SF,x}^{z}z   \notag\\
{s_{\rm x}^{\rm V}(z)}		&=&&~s_{\rm x}^{\rm V,0}+s_{\rm x}^{\rm V,z}z   \notag\\ 
{b_{\rm x}^{\rm 0}(z)}		&=&&
\begin{cases}
~b_{\rm x}^{\rm 0,0}+b_{\rm x}^{\rm 0,z}z  & \text{if $x{=}$`h' or `t',} \notag\\ 
~b_{\rm z}^{\rm 0,0}+b_{\rm z}^{\rm 0,zz}\left(z-z_{\rm b,z}\right)^2  & \text{if $x{=}$`{\rm z}'} \notag
\end{cases}  \\
{b_{\rm x}^{\rm V}(z)}		&=&&~b_{\rm x}^{\rm V,0}+s_{\rm x}^{\rm V,z}z   
\end{alignat}

\noindent  This represents a linear redshift dependence for all of the parameters in Equation \ref{eqn-fit_V_dependence} with the exception of the parameters for the Kaiser boost which we find requires a quadratic dependence for ${b_{\rm x}^{\rm 0}(z)}$ centred on redshift $z_{\rm b,z}$ (hence introducing an extra parameter in this case).  Although very-nearly constant with redshift, we find this refined form of redshift dependence is necessary due to the strong dependence of \sx\ on \bx\ when \Rx\ is only weakly scale dependent (which is always the case for \Rz).

\begin{figure*}
\begin{minipage}{170mm}
\begin{center}
\includegraphics[width=120mm]{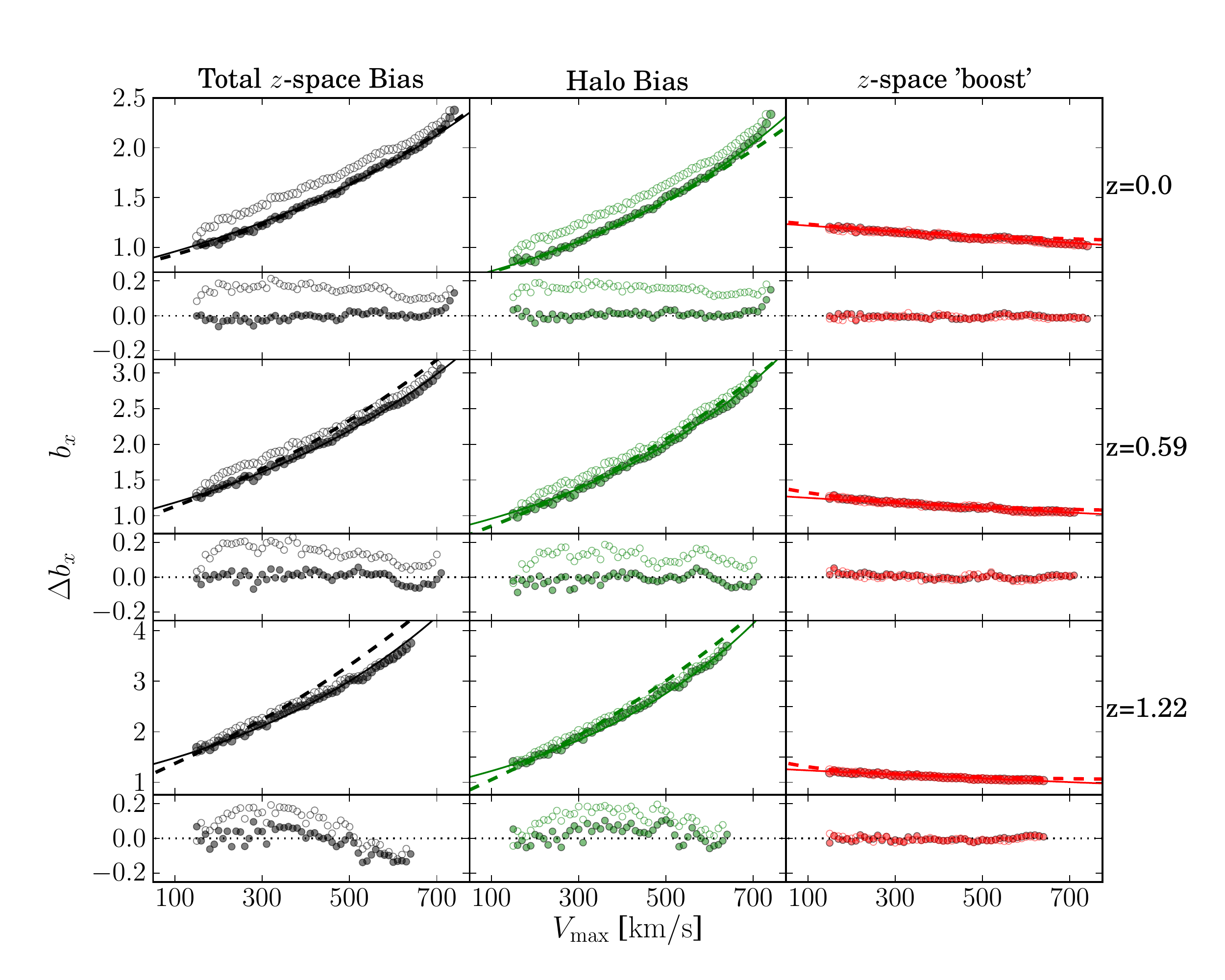}
\includegraphics[width=120mm]{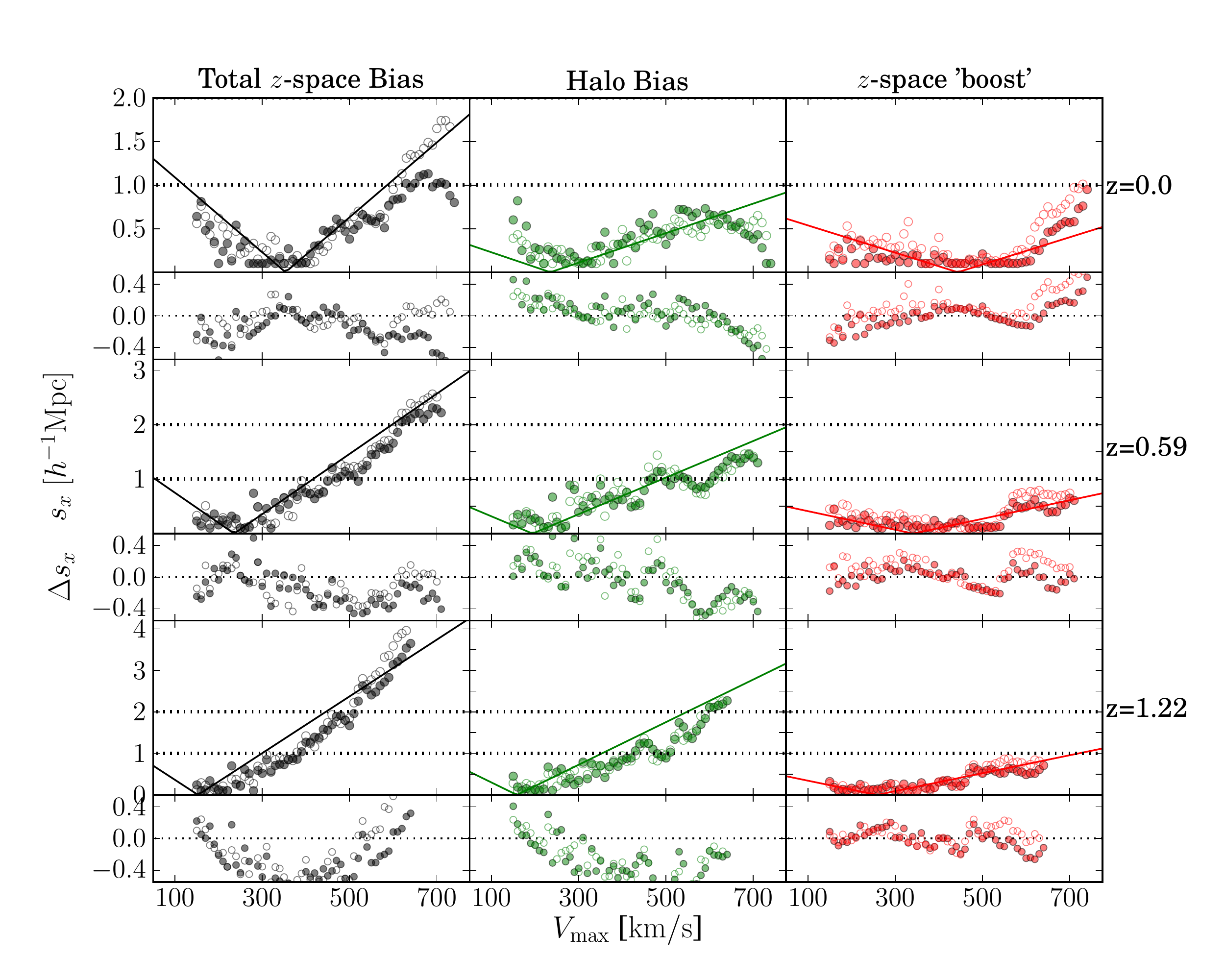}
\caption[effects of substructure] {Plots comparing the large scale bias (top) and amplitude of the scale-dependence of bias (bottom) of friends-of-friends (FoF) halos (solid points) and substructure halos (open points) in the GiggleZ-main simulation.  Solid lines show our final bias model and dashed lines show the simulation-calibrated excursion set model of \citet[][TRK; dashed green]{Tinker:2010}, the redshift-space boost model of \citet[][K87; dashed red]{Kaiser:1987p1760} and the total $z$-space bias resulting from both (dashed black).  Residual differences of each fit from our final global model for the case of FoF halos are plotted as $\Delta s_x$ and $\Delta b_x$ for the large scale bias and amplitude of scale dependant bias, respectively.  Horizontal dotted lines in the right panel denote the values of \sx\ which yield differences in bias between scales $s{=}$3 [$h^{-1}$ Mpc] and $s{=}\infty$ of $15\%$ (\sx{=1} [$h^{-1}$ Mpc]) and $30\%$ (\sx{=2} [$h^{-1}$ Mpc]).}\label{fig-bias_comparison}
\end{center}
\end{minipage}
\end{figure*}

Also presented on Figure \ref{fig-fit_z_dependence} (with lines; solid for FoF halos and dotted for substructure) is the results of our final global MCMC fit to our full dataset.  This fit is applied simultaneously to all of the \Rx\ profiles measured for every grouping at all redshifts employed for this study (and not to the plotted points).  We find that our chosen parameterisation closely follows the individual fits presented with coloured points, validating our assumed form for the redshift dependancies of each parameter.  The resulting parameters describing our full scale-dependent bias model, as described by Equations \ref{eqn-R_of_s} (under the assumption that $\eta{=}1$), \ref{eqn-fit_V_dependence} and \ref{eqn-fit_z_dependence}  are presented in Table \ref{table-fit_results} for both the FoF and substructure halos of our simulation\footnote{A Python script with the full model and its coefficients has been made available online at \url{http://gbpoole.github.io/Poole_2014a_code/}}.  The quality of fit across the whole range of redshifts and masses used to constrain this model are presented in Figure \ref{fig-fit_chi2}.  It is here that the efficacy of our chosen parameterisation should be judged and over the vast majority of the probed mass and redshift range, the quality of fit is very good.  At the highest masses, the quality of fit declines presumably due to overly coarse mass binning demanded by the limited volume available to us for this study.

\subsection{Qualitative trends with mass, redshift and halo type}\label{sec-trends}

Several interesting general trends regarding the dependence of bias (and its scale dependence) on mass, redshift and halo type emerge at this point.  Commenting first on the halo mass at which bias becomes scale free (\VSF), we see that for all bias ratios \VSF\ declines with redshift at a similar rate in all cases and in a nearly identical way for both FoF groups and substructure.  This mass scale is higher for Kaiser boost effects however, leading to a significant increase in this mass scale for the total redshift-space results over that from real-space halo bias effects alone.  For the full redshift range of our study ($z{\lsim}1.2$), \VSF\ is restricted to the range $150[{\rm km/s}]$ to $350[{\rm km/s}]$.  From Equation \ref{eqn-fit_V_dependence} we can see that this trend in \VSFz\ acts to drive an increase in the amplitude of scale dependent bias (\sx) with redshift at masses above this range (where scale dependent bias always results in enhanced bias at small scales) and a suppression of its amplitude on mass scales below it (where scale dependent bias always results in suppressed bias at small scales).

Augmenting these trends in \sx\ driven by the evolution of \VSFz, the mass dependence of \sx\ (given by $s_x^V$) also increases with redshift.  Interestingly, this is the only parameter for which redshift-space contributions to total bias differ between FoF and substructure halos; being significantly higher for substructure, driving an enhanced mass dependence in the total redshift-space bias as well.

We now focus on our results for large-scale bias (\bx).  In Figure \ref{fig-bias_comparison} we illustrate this quantity for all of our groupings at three redshifts spanning the range of our study.  In this case, we directly compare results for FoF (solid points) and substructure halos (open points).  In this figure we also compare our results to the successful simulation-calibrated excursion set model of TRK (dashed green lines), the redshift-space distortion model of K87 (dashed red) and the redshift-space model that emerges from combining the two (dashed black).  The K87 model predicts a redshift-space boost given by (his Equation 3.8, cast here in terms of our notation):
\begin{equation}\label{eqn-Kaiser}
b_z^2{=}1+\frac{2}{3}{\beta}+\frac{1}{5}\beta^{2}
\end{equation}

\noindent where $\beta{=}f/b_h$ with $f$ being the logarithmic derivative of the linear growth factor with respect to expansion factor given by:
\begin{equation}\label{eqn-growth_factor}
f{=}\frac{d\ln D}{d\ln a}
\end{equation}

\noindent Lastly, we also combine the TRK and K87 models to produce a reference total redshift-space bias model (dashed black lines).  

Over most of the range of masses and redshifts probed by our study we find very good agreement between these reference models and our FoF large-scale bias results.  Since the FoF catalogs most straightforwardly relate to the density structures described by excursion set models, this is as expected.  At the highest masses and redshifts, there is a tendency for the TRK model to predict higher real-space biases than our model predicts.  It is possible that the calibration of the TRK model has been biased high from the very strong scale dependant bias of halos in this regime, but this is difficult to discern since their study is conducted in Fourier space and since it is unclear from the presentation of their analysis what exact scale they have fit to.   

Additionally, looking at substructure we find significant enhancements in our large scale real-space (and by extension, total redshift-space) halo biases at low redshift.  This difference is approximately 20\% for Milky Way sized systems (${\sim}220$ km/s) at redshift zero and increases with declining mass.

Interestingly (but perhaps not unexpected), there is absolutely no difference between the two halo populations in terms of their Kaiser boosts.  We interpret this similarity as a reflection of the fact that non-linear pairwise velocities are unimportant on the largest scales of our study.  Furthermore, there is extremely little redshift dependence and only a slight mass dependence for \bz.  We see excellent agreement with the K87 model and interpret the lack of evolution in the Kaiser boost as a remarkable cancelling of the effects on $\beta$ from evolution in the growth of structure (via evolution in $f$) and in real-space halo bias (via evolution in $b_h$).  We note that this level of agreement with the K87 model was also found by \citet[][see their table 4]{Montesano:2010p2505} in their Fourier-space study of bias.

\section{Systematic biases in growth of structure measurements}\label{sec-cosmology_biases}
\begin{figure*}
\begin{minipage}{170mm}
\begin{center}
\vspace{30pt}
\includegraphics[width=170mm]{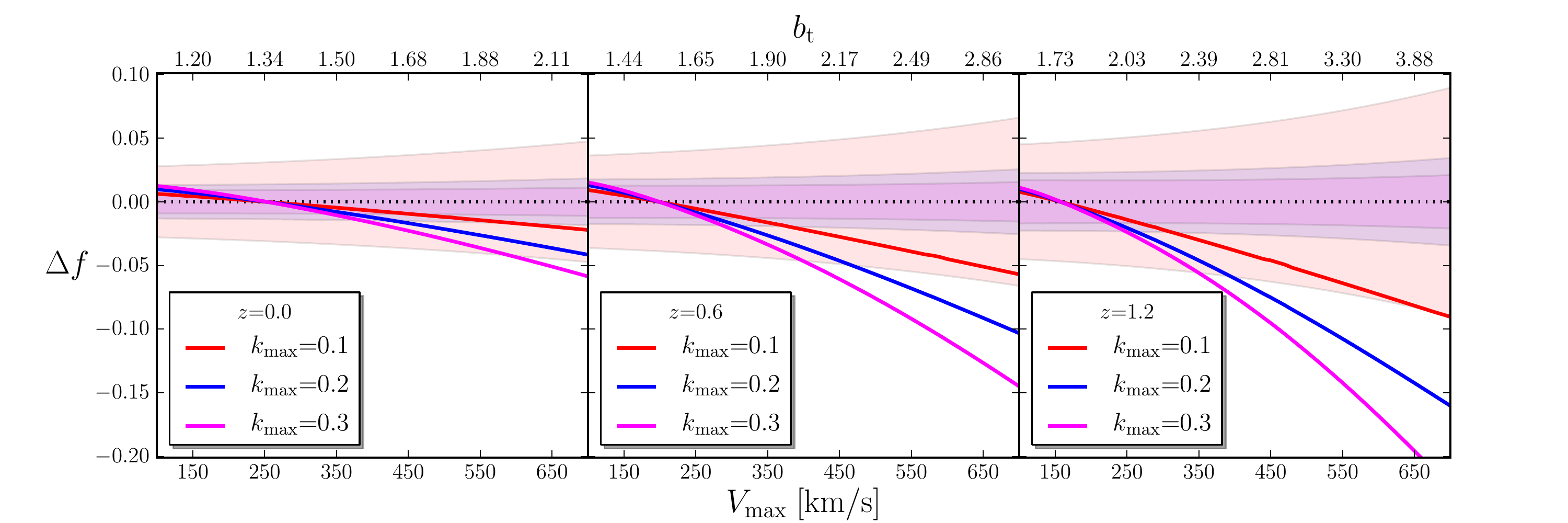}
\caption[Systematic Offsets 1D] {The systematic bias induced in measurements of the growth rate of cosmic structure ($\Delta f$) due to an incorrect assumption of constant galaxy bias ($\Delta f_b$; solid lines) for several cutoff measurement scales ($k_{\rm max}$; for values 0.1,0.2,0.3 $[h^{-1} \rm{Mpc}]^{-1}$ in red, blue and magenta respectively) compared to the statistical uncertainty in this measurement ($\Delta f_s$) for a fiducial survey with volume $1$ $[h^{-1} \rm{Gpc}]^3$ and number density $3{\times}10^5$ $[h^{-1} \rm{Gpc}]^{-3}$ (shaded regions; pink, blue and magenta for $k_{\rm max}{=}$0.1,0.2,0.3 $[h^{-1} \rm{Mpc}]^{-1}$ respectively).\label{fig-systematics_1D}}
\vspace{20pt}
\end{center}
\end{minipage}
\end{figure*}

\begin{figure*}
\begin{minipage}{170mm}
\begin{center}
\includegraphics[width=170mm]{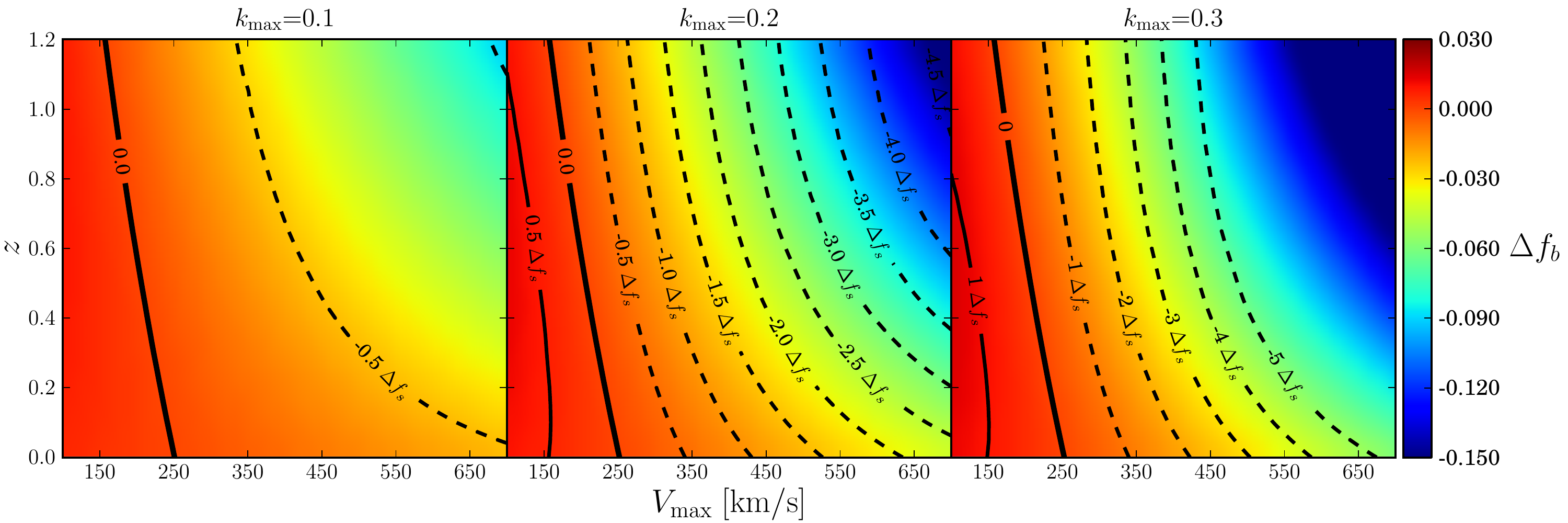}
\includegraphics[width=170mm]{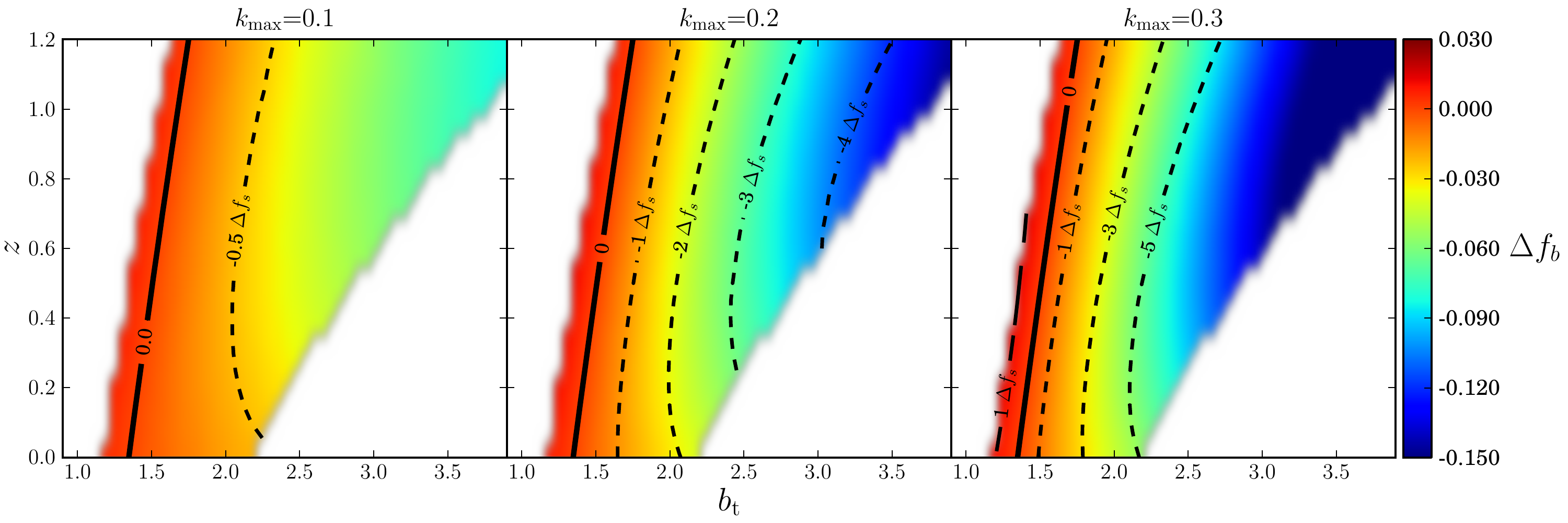}
\caption[Systematic Offsets 2D] {The systematic bias induced in measurements of the growth rate of cosmic structure ($\Delta f_b$) due to an incorrect assumption of constant galaxy bias for several measurement scale cutoffs ($k_{\rm max}$; values given in text above each pannel) as a function of redshift and halo mass (quantified by maximum circular velocity, \Vmax; top) or total redshift-space bias ($b_t$; bottom).  Black contours express this in units of the statistical uncertainty in this measurement ($\Delta f_s$; dashed contours for negative systematic biases, solid lines for positive systematic biases) for a fiducial survey with volume $1$ $[h^{-1} \rm{Gpc}]^3$ and number density $3{\times}10^5$ $[h^{-1} \rm{Gpc}]^{-3}$.  Thick solid contours indicate the cases where scale dependent bias vanishes on scales larger than $3$ $[h^{-1} {\rm Mpc}]$ resulting in no systematic bias in $f$.\label{fig-systematics_2D}}
\end{center}
\end{minipage}
\end{figure*}

Having developed our full parameterisation of scale dependant bias, we seek now to quantify the systematic bias that results in growth of structure measurements when the scale-dependence of bias is not taken into account.  This is done by applying an extension of the Fisher matrix formalism to our bias model in Fourier space where covariance is minimised and measurement uncertainties are more straightforwardly modelled.  We intend for this to be an illustration of the effects of scale dependant bias on measurements of this sort and caution that our estimates here may be somewhat pessimistic.  This is because we will assume a specific and fixed redshift-space distortion model for this calculation whereas fits to data usually marginalise over a velocity-dispersion parameter ($\sigma_v$) which can absorb some of the systematic we present here.  Nevertheless, we expect the general trends and effects presented here to be an informative illustration of the circumstances in which systematic bias should be taken into account in growth of structure studies.

\subsection{Estimation of systematic bias}

\noindent To express our bias model in Fourier space, we first compute an unbiased 2D power spectrum ($P(k,\mu)$, where $\mu{=}\cos(\theta)$ with $\theta$ being the angle between the Fourier mode and the line-of-sight) by applying the K87 redshift-space distortion model to a 1D CAMB power spectrum:
\begin{equation}
P_{\rm model}{(k,\mu)} = (b_h + f \mu^2)^2 P_{\rm CAMB}(k)
\end{equation}

\noindent We then convert this 2D Fourier space model to configuration space using Equation 11 of \citet[][see also \citealt{Hamilton:1998p2546}]{Reid:2012p2511} which relates correlation function multipoles (indexed by $\ell$) to those of its associated power spectrum:
\begin{equation}\label{eqn-P_multipole}
\xi_\ell{(s)} = \frac{i^\ell}{2\pi^2} \int \! P_\ell(k) j_\ell(ks) k^2 \, \mathrm{d}k
\end{equation}

\noindent and apply our bias model to the result.  This is done for both our scale dependent bias model and a constant bias model, yielding (once we convert back to Fourier space) the biased power spectra $P_{\rm model}(k,\mu)$ and $P_{\rm sys}(k,\mu)$ respectively.  

For our estimation of systematic bias in $f$ (which we denote \Dfb) we follow the method of \citet{Amara:2008p2510}.  This method employs a straight-forward extension of the Fisher matrix formalism with the adjustment that uncertainties from systematic biases in parameters are separated explicitly from statistical uncertainties (rather than treated, for example, as an additive term to the statistical contribution to be marginalised over).  Defining the Fisher matrix in the usual way for a 2D power spectrum dependant on a set of parameters $p_i{=}\{f,...\}$ with covariance $C_{ij}$ between each $(k,\mu)$ power spectrum bin as:
\begin{equation}
F_{ij} = \sum_{(k,\mu)} C^{-2}_{ij} \frac{dP_{\rm model}(k,\mu,p)}{dp_i} \frac{dP_{\rm model}(k,\mu,p)}{dp_j}
\end{equation}

\noindent the systematic uncertainty in each parameter (which we denote generically as $\Delta p_i$) is obtained by projecting the inverse of this Fisher matrix along a bias vector $B_{j}$ as given by the expression
\begin{equation}
\Delta p_i = F^{-1}_{ij} B_j
\end{equation}

For the parameter $p_0{=}f$, the relevant expression for $B_{0}$ is found by rewriting Equation 8 of \citet{Amara:2008p2510} in the following form:
\begin{equation}
B_0 = \sum_{(k,\mu)} \frac{\Delta P_{\rm sys}(k,\mu,f)}{\sigma^2_P(k,\mu)}\frac{dP_{\rm model}}{df}(k,\mu,f)
\end{equation}

\noindent with $\sigma_P(k,\mu)$ being the error in each 2D power spectrum bin given by:
\begin{equation}
\sigma_P(k,\mu) = \frac{P(k,\mu) + 1/n}{\sqrt{N}}
\end{equation}

\noindent with $n$ being the number density of galaxies and $N$ the number of Fourier modes in each bin.  The quantity $\Delta P_{sys}{=}P_{\rm model}{-}P_{\rm sys}$ represents the residual systematic modelling error in the power spectrum.  Lastly, $dP_{model}/df$ gives the partial derivative of our model power spectrum with respect to $f$.  Throughout, we use bin widths  of $\Delta k{=}0.01$ $[h^{-1} \rm{Mpc}]^{-1}$ and $\Delta \mu{=}0.1$ for sums over $k$ and $\mu$ respectively.

\subsection{Effects of systematic bias}

To evaluate the magnitude of this systematic bias, we express it here for a fiducial survey of volume $1$ $[h^{-1} \rm{Gpc}]^3$ and number density $n{=}3{\times}10^5$ $[h^{-1} \rm{Gpc}]^{-3}$.  This number density is chosen to be similar to that of both the WiggleZ and BOSS Surveys and the volume is representative of current large spectroscopic surveys.  In Figure \ref{fig-systematics_1D} we show the results of this calculation at three redshifts spanning the range $z{\lsim}1.2$ for three small-scale cutoffs (denoted $k_{\rm max}$).  These are compared in each case to the statistical uncertainty expected for this measurement (denoted $\Delta f_s$; shown with shaded regions) which we calculate using a standard Fisher matrix forecast \citep[see][]{White:2009p2529,Abramo:2012p2530,Blake:2013p2531} using the same binning and range as for the systematics forecast.

Noting first some generic trends in this figure, we see that \Dfb\ is positive for low masses/biases and (more generally) negative for larger masses/biases.  This is due to the transition from $\mathcal{S}{=}{-}1$ (suppression of bias on small scales) to $\mathcal{S}{=}{+}1$ (enhancement of bias on small scales) with suppressed small-scale bias leading to a positive bias in $f$ and enhanced small-scale bias (the more common case) leading to a negative bias in $f$.  Additionally, we see that increasing \kmax\ has two distinct effects: it increases the precision of the measurement (particularly between \kmax${=}0.1$ and \kmax${=}0.2$ $[h^{-1} \rm{Mpc}]^{-1}$) due to the additional data involved and it increases \Dfb\ due to the use of scales where scale-dependent bias has an increased effect on the shape of the power spectrum.

Commenting more specifically, we can see from this figure that when \kmax${=}0.1$ $[h^{-1} \rm{Mpc}]^{-1}$, \Dfb\ remains significantly smaller than \Dfs\ for all cases with $b_t{\lsim}2$.  Indeed, only when $b_t{\gtrsim}3$ at $z{\gtrsim}1$ does the systematic bias become significant compared to the precision of the measurement.  However, this situation dramatically changes for larger values of \kmax.  When it increases to $0.2$, \Dfb\ becomes significant compared to \Dfs\ for all cases except those very narrowly similar in mass to $V_{\rm SF}$, where scale dependent bias disappears.

The presentation of these results is expanded in Figure \ref{fig-systematics_2D} where we show \Dfb\ for the full range of cases to which our bias model has been constrained.  Across all redshifts and for all cases, we see that scale dependant bias effects are minimised when the halo population has a bias similar to $b_t{\sim}1.5$.

\section{Summary and conclusions}\label{sec-summary}
We have used the GiggleZ-main simulation to produce an 8-parameter phenomenological model quantifying halo bias (in both real and redshift-spaces) and its scale dependence over the range of masses $100[{\rm km/s}] {<}{V_{\rm max}}{<}700[{\rm km/s}]$, redshifts $z{\lsim}1.2$ and scales $3[{\rm Mpc/h}] {<}{s}{<}100[{\rm Mpc/h}]$ under the ansatz that bias converges to a scale independent form at large scales.  We find that scale dependent bias can either enhance or suppress bias at small scales.  For any given halo mass at any given redshift, large-scale bias is given by a single constant and the scale dependence of bias is given by two others: a binary parameter determining whether bias is enhanced or supressed on small scales ($\mathcal{S}$) and a parameter setting its amplitude ($s$).

While a relatively small but growing body of literature has looked at scale dependent bias effects in the Fourier domain, few recent studies have addressed it in configuration space.  The results presented in this work should not only be more directly applicable to observational studies conducted in configuration space, but should also help provide a basis upon which to build some intuition regarding the scale-dependent bias effects observed in Fourier-space studies.

We find several interesting trends (noted and discussed in Section \ref{sec-trends}) which require further study to understand.  Most prominent among these is the fact that scale dependence of bias transitions from the suppression of bias at small scales for small masses to enhancement for large masses.  It does so in a narrow bias range centred on $b_t{\sim}1.5$ across all redshifts $z{\lsim}1.2$.  At the transition between, our parameterization describes bias as scale free.  It is important to note however, that while a scale free model is a good fit in these regimes, the claim of a scale-free bias can only be made to a precision allowed by our simulation.  Additionally, we wish to emphasise that the functional form of our parameterisation as well as the specific parameter values we have obtained may need to vary under reasonable changes from our fiducial cosmology.  Further study is needed to determine how sensitively they do so. 

It should also be noted that we restrict our study to configuration space on scales larger than $3 [{\rm Mpc/h}]$.  On scales lower than this, a wide variety of non-monotonic variations in \Rx\ occur \citep[of a character similar to that presented in figure 4 of][]{Zehavi:2004p2535}.  In the Fourier domain, these features are likely to have broad spectral content and more detailed study is required to understand their influence in Fourier space.

Lastly, we compute the systematic biases induced in growth of structure measurements in the absence of corrections for scale-dependent bias effects.  We find that for a fiducial survey with volume $1$ $[h^{-1} \rm{Gpc}]^3$ and number density $n{=}3{\times}10^5$ $[h^{-1} \rm{Gpc}]^{-3}$ that systematic bias is modest when scales only as small as $k_{\rm max}{=}0.1$ are used, except for highly biased halos at high redshift.  Once scales as short as $k_{\rm max}{\gtrsim}0.2$ are utilised, the situation dramatically changes with significant systematic biases resulting at all redshifts for biases even just slightly different from $b_t{\sim}1.5$.  In realistic analysis where fits are generally marginalised over a pair-wise velocity dispersion parameter, much of this effect is likely to be absorbed into this parameter, reducing the problem at the expense of compromising any meaning given to this quantity.  Further study under realistic conditions is clearly needed to precisely quantify these effects on real survey results.

These results suggest that the optimal strategy at all redshifts $z{\lsim}1.2$ for clustering studies which are dominated more by systematic effects than statistical precision (such as the case of cosmological neutrino mass measurements) is to target $b_t{\sim}1.5$ systems.  Fortuitously, the UV-selected galaxies targeted by the WiggleZ survey have a large-scale bias similar to this \citep{Blake:2009p75} for example.

These results reenforce the notion that scale dependent bias is particularly significant for studies involving measurements of the $\emph{shape}$ of two-point clustering statistics.  We have focused here on growth of structure measurements only, but similar analysis \citep[following-on from the work of][for example]{Swanson:2010p2539} for neutrino mass  measurements are clearly warranted as well.

Of course, this study has focused on the bias properties of halo tracers with complete selection properties and uniform masses.  The larger bias we find for substructure catalogs shows the importance of realistically considering the sites of galaxy formation.  We now need to carefully consider the effects that can be induced by the sorts of colour selections employed during observational campaigns.  Due to phenomena like the morphology-density relation, a large variety of differing results can occur if galaxies are selected by more observationally motivated criteria (\eg\ luminosity or colour) which are more difficult to robustly model.  A great deal more study on these issues is required to make robust statements under such circumstances.

\section*{Acknowledgements}
We would like to thank Volker Springel for making GADGET-2 publicly available and for permitting us to use his halo finding code (Subfind) and Aaron Ludlow for useful comments.  We would also like to thank Swinburne University for its generous allocation of computing time (on the Green and gSTAR facilities)  for this project, Jarrod Hurley and the Green Machine Help Desk for their computing support, and the Green Machine user community for patiently waiting while the GiggleZ-main simulation kept them from their work.  We acknowledge financial support from the Australian Research Council through Discovery Project grants DP0772084 and DP1093738.  GP and SM acknowledge support from the ARC Laureate program of Stuart Wyithe.  CB and CP acknowledges the support of the Australian Research Council through the award of a Future Fellowship.  DC acknowledge the support of an Australian Research Council QEII Fellowship.

\bibliographystyle{mn2e}
\bibliography{biblio}

\end{document}